\documentclass[a4paper]{aa}

\usepackage{graphicx,natbib,amssymb,url,txfonts,hyperref} 
\bibpunct{(}{)}{;}{a}{}{,}

\graphicspath{{figures/}}
  
\newcommand{\solarlum}{L_{\sun}}  
\newcommand{\solarmass}{M_{\sun}}

\newcommand{\cmsqr}{\mathrm{cm}^{-2}}
\newcommand{\cmcub}{\mathrm{cm}^{-3}}
\newcommand{\kms}{\mathrm{km} \, \mathrm{s}^{-1}}
\newcommand{\kkms}{\mathrm{K} \, \kms}

\newcommand{\Xout}{X_\mathrm{out}}
\newcommand{\Xin}{X_\mathrm{in}}

\newcommand{\cf}{\emph{cf.}\ }
\newcommand{\eg}{\emph{e.g.}\ }

\newcommand{\tabnote}[1]{$^\mathrm{#1}$}

\defcitealias{Maret04}{Paper~I}

\begin{document} 
 
\date{Received <date> / Accepted <date>} 

\title{CH$_\mathsf{3}$OH abundance in low mass protostars}

\titlerunning{CH$_\mathsf{3}$OH in low mass protostars}
 
\author{S. Maret\inst{1}\thanks{\emph{Present address:} University of
    Michigan, Department of Astronomy, 500 Church St., Ann
    Arbor MI 48109-1042, USA} \and C. Ceccarelli\inst{1} \and
    A.~G.~G.~M. Tielens\inst{2} \and E. Caux\inst{3} \and
    B. Lefloch\inst{1} \and A. Faure\inst{1} \and A. Castets\inst{4}
    \and D.~R. Flower\inst{5} }
 
\institute{Laboratoire d'Astrophysique, Observatoire de Grenoble, BP
   53, F-38041 Grenoble Cedex 09, France \and Space Research
   Organization of the Netherlands, PO Box 800, 9700 AV Groningen,
   The Netherlands \and Centre d'Etude Spatiale des Rayonnements,
   CESR/CNRS-UPS, BP 4346, F-31028 Toulouse Cedex 04, France \and
   Observatoire de Bordeaux, BP 79, F-33270 Floirac, France \and Physics
   Department, The University, Durham DH1 3LE, UK
}
 
\offprints{S\'ebastien Maret, \email{smaret@umich.edu}}

\date{Received / Accepted} 

\abstract{We present observations of methanol lines in a sample of
  Class 0 low mass protostars. Using a 1-D radiative transfer model,
  we derive the abundances in the envelopes. In two sources of the
  sample, the observations can only be reproduced by the model if the
  methanol abundance is enhanced by about two order of magnitude in
  the inner hot region of the envelope. Two other sources show similar
  jumps, although at a lower confidence level. The observations for
  the other three sources are well reproduced with a constant
  abundance, but the presence of a jump cannot be ruled out. The
  observed methanol abundances in the warm gas around low mass
  protostars are orders of magnitude higher than gas phase chemistry
  models predict. Hence, in agreement with other evidence, this
  suggests that the high methanol abundance reflects recent
  evaporation of ices due to the heating by the newly formed star.
  The observed abundance ratios of CH$_3$OH, H$_2$CO and CO are in
  good agreement with grain surface chemistry models. However, the
  absolute abundances are more difficult to reproduce and may indicate
  the presence of multiple ice components in these regions.

}

\maketitle
     
\section{Introduction}
\label{sec:introduction}

During the formation of a star, the gas undergoes important physical
and chemical changes. In the prestellar phase, the gas is heavily
depleted by accretion on grain mantles. When the gravitational
collapse starts, the protostar warms the gas while ice mantle
molecules are released into the gas phase. These released molecules
trigger the formation of more complex molecules through rapid
reactions in the warm gas \citep{Charnley92}.

While the importance of ice mantle evaporation and hot core chemistry
around high mass protostars was known well over a decade ago, the
existence of such regions in low mass protostars has been established
more recently. \citet{Ceccarelli00a, Ceccarelli00b} have shown that
H$_2$O and H$_2$CO abundances in the low mass protostar IRAS16293-2422
are increased in the warm and dense inner part of the circumstellar
envelope. In this region, H$_2$O and H$_2$CO are evaporated from the
grain mantles and are injected in the gaseous phase. These findings
were later confirmed by \citet{Schoier02}, who found strong evidence
for the increase of several other molecular abundances (e.g. CH$_3$OH,
SO and SO$_2$). Evaporation of H$_2$O has also been shown around one
other Class 0 protostar, NGC1333-IRAS4 \citep{Maret02}. Recently,
\citet[][ Paper I]{Maret04} carried out a survey of the emission of
H$_2$CO in a sample of nine low mass protostars, and concluded that in
all the observed protostars but one, H$_2$CO is between two and three
orders of magnitude more abundant in the inner part of the envelope
than in the outer part, suggesting that the evaporation of grain
mantles is a common phenomenon in the inner parts of low mass
protostars.

While the importance of grain mantle evaporation is now well
established in low mass protostars, the question of whether these
molecules will form ``daughter'' molecules by \emph{hot core}
chemistry \citep{Charnley92} is still open. From a theoretical point
of view, if the protostar is undergoing collapse, only very rapidly
formed second-generation complex molecules can be produced. As noted
by \citet{Schoier02} if the protostar is in free fall, the evaporated
molecules will fall on the central object in a few hundreds
years. Hence, while evaporation can still occur on such a rapid
timescale, the composition of the gas phase is then `frozen' in,
because the dynamical time scale is much shorter than the time needed
to form second generation molecules \citep{Rodgers03}. However,
observationally, the recent detections of complex O and N bearing
molecules -- typical of massive hot cores -- towards IRAS162293-2422
\citep{Cazaux03,Bottinelli04b} and NGC1333-IRAS4A
\citep{Bottinelli04a} are challenging these theoretical concepts,
emphasising that the chemistry in low mass protostar envelopes is
still poorly understood and needs to be further investigated.

Methanol is particularly suited for the study of hot cores around low
mass protostars, because it is relatively abundant in grain mantles
and dominates the organic content of evaporation zones. Indeed, such
enrichments of methanol have already been observed in the inner
regions of high mass protostars \citep{vanderTak00b}, and in the gas
shocked by prostellar outflows \citep{Bachiller98,
Buckle02}. Moreover, methanol, being a slightly asymmetric-top
molecule, can be used to probe a large range of physical conditions,
both in density and temperature \citep[e.g.][]{Leurini04}. The recent
computations of collisional rate coefficients of methanol with
para-H$_2$ \citep{Pottage04} make precise determination of these
conditions possible.

In this paper, observations of methanol transitions from low mass
protostars are presented. A detailed radiative transfer model is used
to derive the methanol abundance inside the protostellar
envelopes. Finally, the methanol abundances in the envelopes are
compared to those in other environments, and the implications for the
formation of this molecule are discussed.

\section{Observations}
\label{sec:observations}

Six Class 0 protostars (see Table \ref{tab:sources}) were observed
using the Institut de Radio-Astronomie Millim\'etrique 30 meters
telescope (IRAM-30m)\footnote{IRAM is an international venture
supported by INSU/CNRS (France), MPG (Germany) and IGN (Spain).}, and
the James Clerk Maxwell Telescope (JCMT)\footnote{The JCMT is operated
by the Joint Astronomy Center in Hilo, Hawaii on behalf of the present
organizations: The Particle Physics and Astronomy Research Council in
the United Kingdom, the National Research Council of Canada and the
Netherlands Organization for Scientific Research.}. The methanol
5$_K$-4$_K$ transitions were observed with the IRAM-30m telescope,
while the 7$_K$-6$_K$ were observed with the JCMT. JCMT and CSO
observations of IRAS16293-2422 from the literature were also used
\citep{vanDishoeck95}.

\begin{table*}
  \caption{The observed sample.}
  \begin{center}
    \begin{tabular}{l l l l l l l l l}
      \hline
      \hline
      Source & $\alpha (2000)$ & $\delta (2000)$ & Cloud &
      Dist.\tabnote{a} & $L_\mathrm{bol}$\tabnote{b} &
      $M_\mathrm{env}$\tabnote{b} & 
      $L_\mathrm{smm} / L_\mathrm{bol}$\tabnote{c} & 
      $T_\mathrm{bol}$\tabnote{c} \\ &&&&
      (pc) & ($\solarlum$) & ($\solarmass$) & (\%) & (K)\\
      \hline
      NGC1333-IRAS4A & 03:29:10.3 & +31:13:31 & Perseus & 220
      & 6 & 2.3 & 5 & 34\\
      NGC1333-IRAS4B & 03:29:12.0 & +31:13:09 & Perseus & 220
      & 6 & 2.0 & 3 & 36\\
      NGC1333-IRAS2 & 03:28:55.4 & +31:14:35 & Perseus & 220
      & 16 & 1.7 & $\lesssim$ 1 & 50\\
      L1448-MM & 03:25:38.8 & +30:44:05 & Perseus & 220 & 5 &
      0.9 & 2 & 60\\
      L1448-N & 03:25:36.3 & +30:45:15 & Perseus & 220 & 6 &
      3.5 & 3 & 55\\
      L1157-MM & 20:39:06.2 & +68:02:22 & Isolated & 325 & 11
      & 1.6 & 5 & 60\\
      IRAS16293-2422 & 16:32:22.7 & -24:38:32 &
      $\rho$-Ophiuchus & 160 & 27 & 5.4 &  2 & 43\\
      \hline
    \end{tabular}
  \end{center}
  \tabnote{a} From \citet{Andre00}, except for Perseus
  sources \citep{Cernis90}. \\
  \tabnote{b} From \citet{Jorgensen02}. \\
  \tabnote{c} From \citet{Andre00}. \\
  \label{tab:sources}
\end{table*}

The IRAM-30m observations were obtained in November 2002. The A230 and
B230 receivers were used simultaneously with the autocorrelator,
providing a spectral resolution of 0.5 km/s and a bandwidth of 320
MHz.  The calibration and the pointing were regularly checked and were
found to be better than 20\% and 3\arcsec \ respectively. For all the
sources the observations were done in position switching mode, after
checking that the reference position was free of line emission. In
order to determine the spatial extent of the methanol emission, small
maps of 3 by 3 pixels were made, with a 10\arcsec \ sampling.

The JCMT observations were obtained in September 2000 and from
February 2001 to February 2003. The B3 receiver was used with the
digital autocorrelator spectrometer (DAS) in setup with a bandwith of
500 MHz and a spectral resolution of 0.4 km/s. As for the IRAM-30m
observations, the pointing and calibration were regularly checked and
was found to be better than 30\% and 3\arcsec \ respectively. A few
lines already observed by \citet{vanDishoeck95} towards IRAS16293-2422
were re-observed in September 2000. The lines fluxes from the two
datasets were compared and were found to be in agreement within
30\%. The JCMT observations were done in beam switching mode with a
180\arcsec\ offset, large enough to avoid line contamination by the
outflows or by the cloud.

\section{Results}
\label{sec:results}

\begin{figure*} 
  \centering 
  \includegraphics[width=17cm]{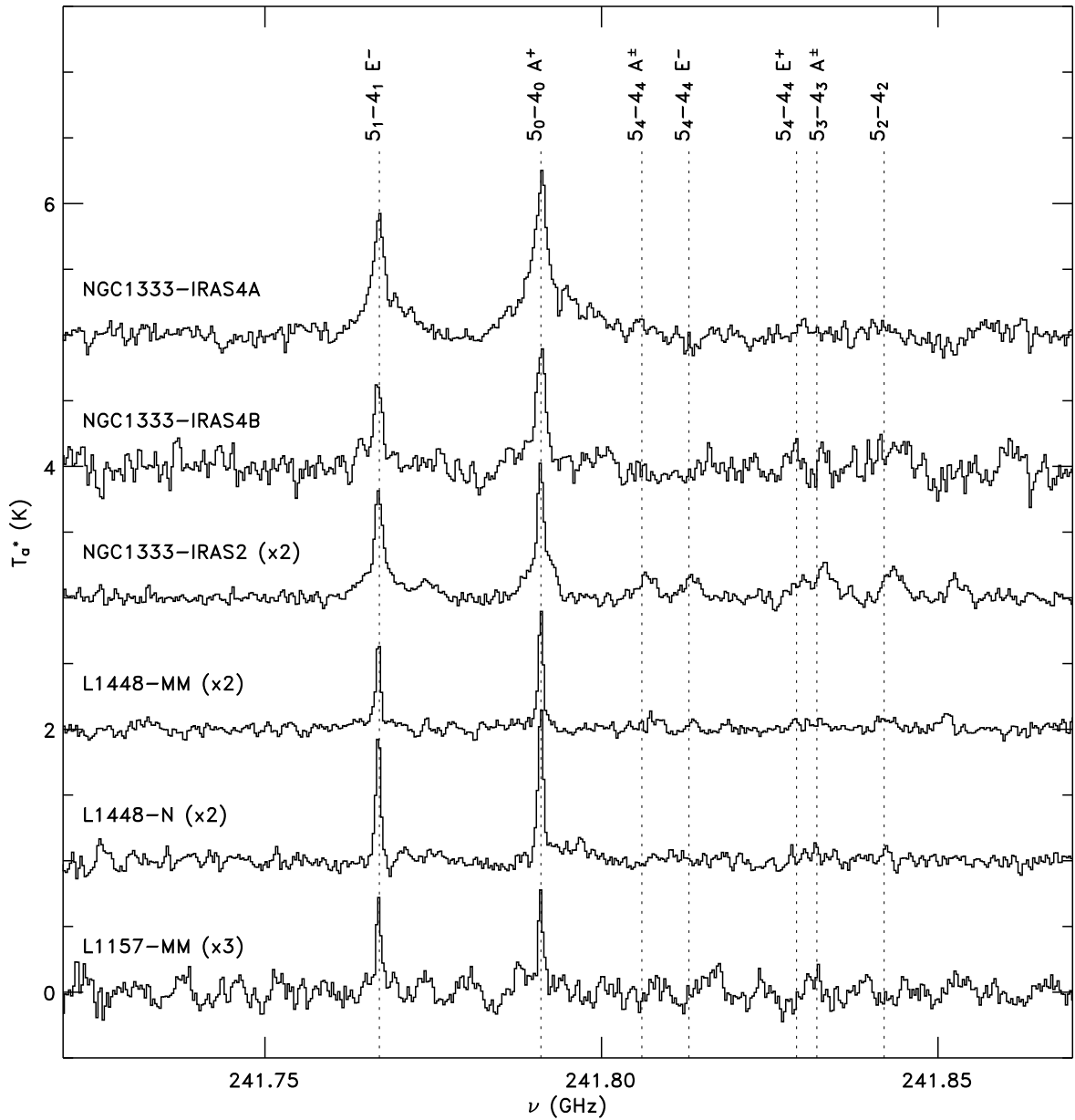} 
  \caption{On source CH$_3$OH 5$_K$-4$_K$ spectra.}
\label{fig:spectra_1}
\end{figure*}

\begin{figure*} 
  \centering 
  \includegraphics[width=17cm]{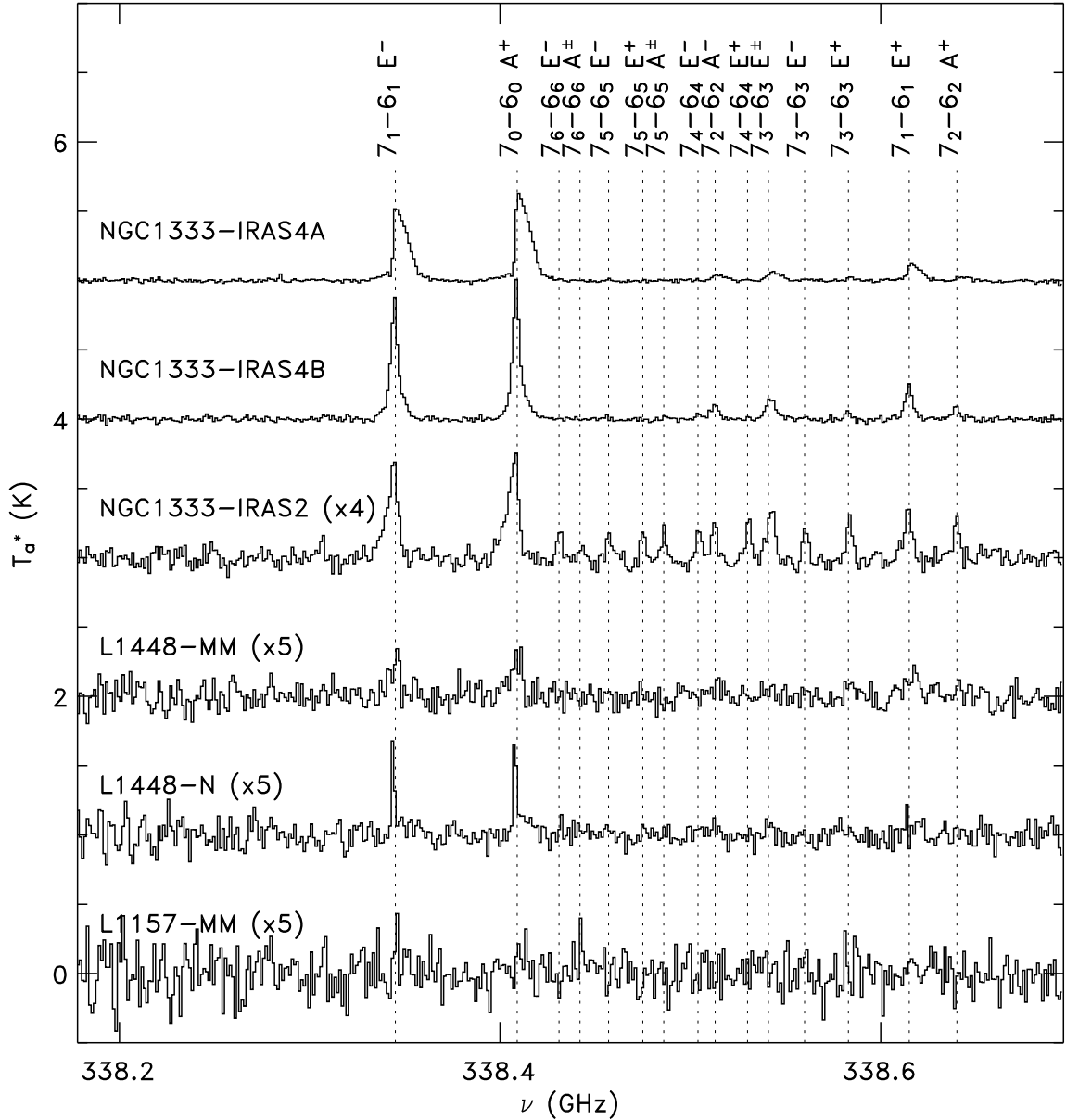} 
  \caption{As in Fig. \ref{fig:spectra_1} for CH$_3$OH 7$_K$-6$_K$.}
  \label{fig:spectra_2} 
\end{figure*}

Fig. \ref{fig:spectra_1} and \ref{fig:spectra_2} show the spectra
towards the six observed sources. Several 5-4 and 7-6 lines are
detected in the bands observed with IRAM and JCMT respectively. The
lines with lowest level energy (E$_{\mathrm{up}}$ $\sim$ 40 - 70 K) in
each band are the brightest in the spectra and are detected toward all
sources. On the contrary, only a few lines with larger upper level
energies are detected in most sources, with NGC1333-IRAS2 showing the
largest number of detected methanol lines (26 lines, up to upper level
energies of 259 K). The relative intensity of the lines with low level
energy to lines with high level energy varies from source to source,
suggesting different excitation temperatures. The richer methanol
spectrum of NGC1333-IRAS2 is, in this respect, likely due to a
relatively larger excitation temperature with respect to other
sources.

The lines with lowest level energy show, in some sources, bright high
velocity wings, sometimes asymmetric. This is the case for
NGC1333-IRAS4A, NGC1333-IRAS4B and NGC1333-IRAS2. In the other three
sources (L1448-MM, L1448-N and L1557-MM), all the lines, including
those with a low level energy, have relatively narrow Gaussian
profiles (between 1 and 4 $\kms$) with much weaker broad wings, if
any. Usually, large asymmetric wings testify to the presence of
outflowing material.  Indeed, the maps obtained at IRAM in the
direction of NGC1333-IRAS4A and NGC1333-IRAS2 confirm the presence of
large scale outflows probed by the two methanol lines with lowest
level energy.

\begin{figure*} 
  \centering 
  \includegraphics[width=17cm]{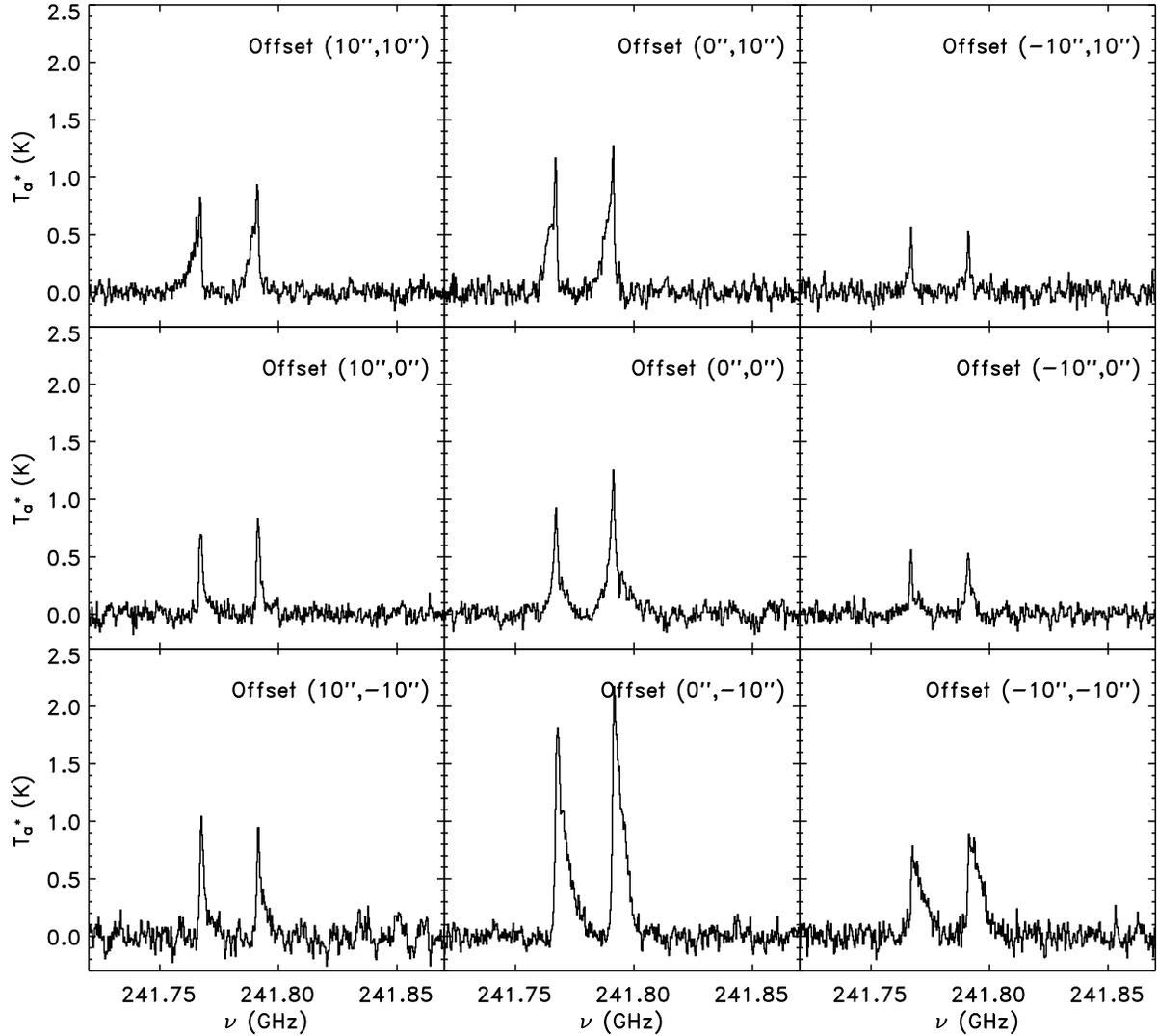} 
  \caption{CH$_3$OH 5$_K$-4$_K$ maps of NGC1333-IRAS4A. The outflow is
    elongating on a north south axis (see text).} 
  \label{fig:IRAS4A_map} 
\end{figure*}

Fig. \ref{fig:IRAS4A_map} shows the CH$_3$OH 5$_K$-4$_K$ emission map
of NGC1333-IRAS4A. Towards the central position, the two lines with
lowest level energy have Gaussian profiles with high velocity
wings. North and south of the source, the wings become larger and
asymmetric, red to the north-east and blue to the south-west. Also,
the emission of the two lines with lowest level energy is not peaked
in the central position, but rather 10\arcsec \ south of it. The
morphology of the observed emission is consistent with the direction
of the outflow elongating along the north-south axis near
NGC1333-IRAS4A\footnote{There is an abrupt change of the position
angle of the CO 3-2 flow from approximatively 45$\degr$ \ on the large
scale to 0$\degr$ \ near NGC1333-IRAS4A \citep{Blake95}.}, as seen in
CO 3-2, CS 7-6 or SiO 2-1 emission \citep{Blake95, Lefloch98}.

\begin{figure*} 
  \centering 
  \includegraphics[width=17cm]{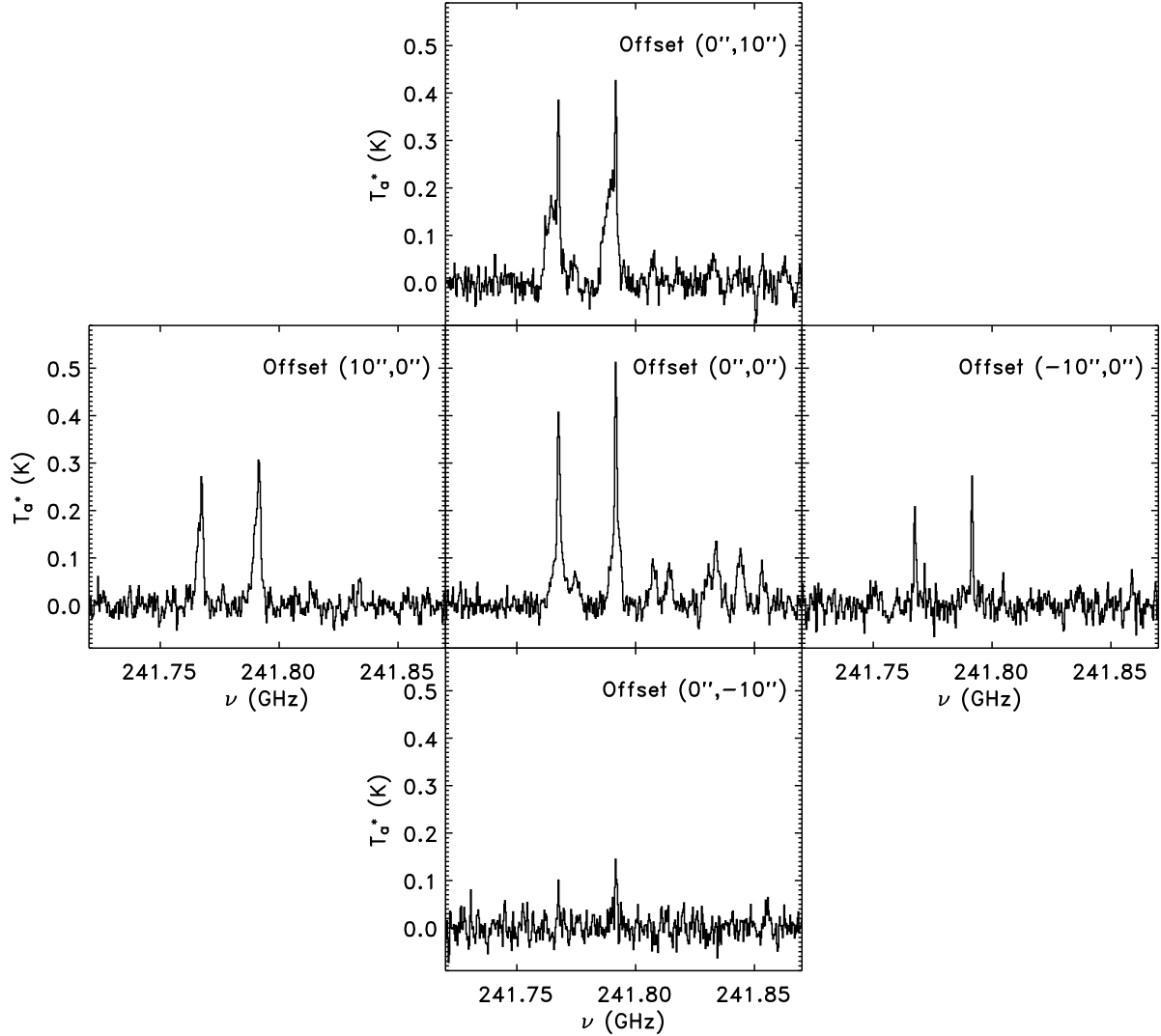} 
  \caption{As in Fig. \ref{fig:IRAS4A_map} for NGC1333-IRAS2. A first
    outflow is oriented towards north-south, and a second one towards
    east-west (see text).}
  \label{fig:IRAS2_map} 
\end{figure*}

The map of CH$_3$OH 5$_K$-4$_K$ emission of NGC1333-IRAS2 is presented
in Fig. \ref{fig:IRAS2_map}. In the central position, the lines
profiles are similar to those observed in NGC1333-IRAS4A: a narrow
Gaussian component, with high velocity wings, although less intense.
The emission becomes red-shifted north and eastward of the source, and
decreases in the west and south direction. Two outflows have been
observed towards NGC1333-IRAS2. A large scale outflow, oriented along
the north-south axis of the source, has been detected in CO 3-2 and
2-1 emission \citep{Knee00}.  A second outflow, oriented east and
west, is detected in SiO 2-1 \citep{Blake96, Jorgensen04b} and CS 3-2
\citep{Langer96}. Strong CH$_3$OH lines have also been observed at the
endpoint of this outflow ($\sim$ 70\arcsec \ east and west of the
central position), where the methanol abundance is enhanced by a
factor of 300 \citep{Bachiller98}. In our map, low energy transitions
trace the red lobe (towards the north) of the outflow, as well as the
blue (towards the west) lobe of the east-west outflow, but only narrow
and weak lines are seen towards the east and south. On the contrary,
lines with high energy, also seen on the JCMT spectrum, appear only in
the central position.

Since this paper focuses on the methanol emission from the envelopes
surrounding the protostars, in the following we will restrain the
discussion to this component only and will not analyze further the
outflow component, other than for disentangling it from the envelope
emission. In practice, we separated the envelope emission, assumed to
give rise to a Gaussian line centered on the v$_{\rm lsr}$ of the
source, from the outflow emission, assumed to be the residual of the
Gaussian. Tables \ref{tab:flux_iras4a} to \ref{tab:flux_iras16293}
list the line fluxes and widths of the Gaussians, for each source, as
derived from this analysis. As previously said, in most cases the
lines are rather ``clean'' Gaussians with narrow widths, and little
contamination from wings. However, a few lines are strongly
contaminated by the outflow component, and such a separation between
the envelope and outflow components was not possible. In these cases,
a single Gaussian was fitted on the entire profile, therefore
including both envelope and outflow contributions. These lines are
explicitly marked in Tables \ref{tab:flux_iras4a} to
\ref{tab:flux_iras16293}.

\begin{figure*}
  \centering
  \includegraphics[width=17cm]{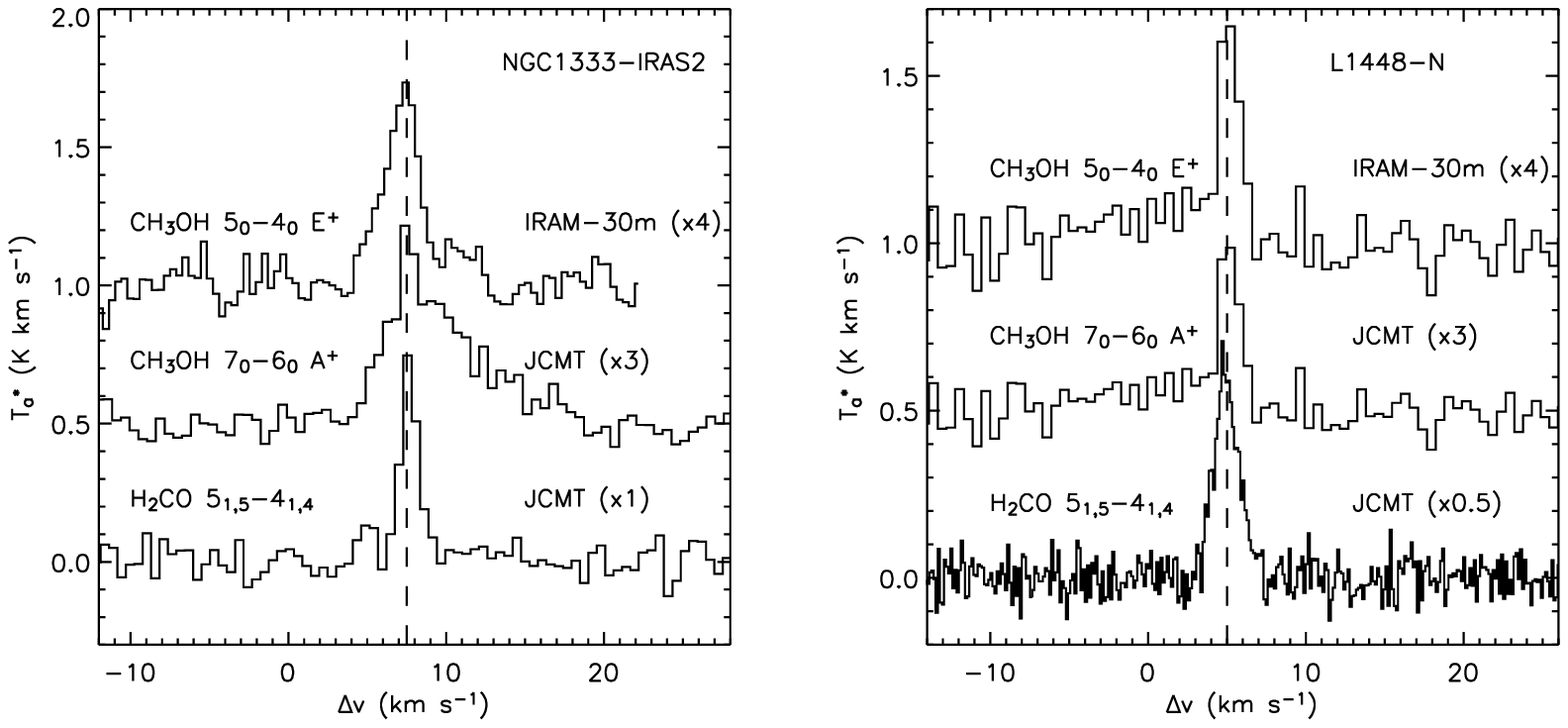}
  \caption{Comparison between two methanol (this paper) and one
    formaldehyde line profiles \citepalias{Maret04} from NGC1333-IRAS2
    and L1448-N. The vertical dashed line indicates the
    $v_\mathrm{lsr}$ of the sources.}
  \label{fig:IRAS2_lines}
\end{figure*}

In the three sources where the outflow emission is relatively
important (NGC1333-IRAS4A, IRAS4B and IRAS2), several lines observed
with IRAM are significatively narrower than the lines observed with
the JCMT: between 1 and 4 $\kms$ the formers, and up to 8 $\kms$ the
latters (before outflow component subtraction).
Fig. \ref{fig:IRAS2_lines} illustrates this difference, and presents
two methanol line profiles observed with IRAM-30m and JCMT towards
NGC1333-IRAS2.  The line observed with the JCMT clearly shows a red
high velocity wing, while the IRAM spectrum does not. Likely, this
difference is due to the larger beam of JCMT (13\arcsec) with respect
to IRAM-30m (10\arcsec).  The former probably picks up more extended
emission from the outflows than the latter. In addition, or
alternatively, this difference can be due to a slightly different
pointing of the two telescopes.

The second panel of Fig. \ref{fig:IRAS2_lines} shows the case of
L1448-N, where, on the contrary, the lines from the two telescopes are
very similar, confirming the lower contribution from the outflow to
the line emission in this source. Fig. \ref{fig:IRAS2_lines} also show
a formaldehyde line along with the two methanol IRAM and JCMT lines.
The figure shows that, in the case of NGC1333-IRAS2, the JCMT methanol
line is indeed much more contaminated by the outflow than the
formaldehyde line, also obtained at JCMT.  However, inspection of the
spectra obtained towards L1448-N does not show any substantial
difference, supporting our conclusion that, in this source, the
outflow does not play a major role in the methanol emission.

In summary, profiles and maps of the detected methanol lines suggest
that in half of the sample sources (NGC1333-IRAS4A, NGC1333-IRAS4B and
NGC1333-IRAS2) the lines with lowest level energy are strongly
contaminated by the outflowing gas, whereas in the the lines with the
highest level energy the problem is much less severe.  In the
remaining three sources (L1448-MM, L1448-N and L1157-MM) no evidence
of severe outflow contamination has been observed in any methanol
line. For most of the lines showing contamination by the outflow, the
outflow emission was disentangled from the envelope emission. For a
few lines marked in Tables \ref{tab:flux_iras4a} to
\ref{tab:flux_iras2}, such a separation was not possible.  In all
other cases, we are confident that the quoted fluxes are
representative of the emission from the envelope only.

\section{Modeling}
\label{sec:modeling}

\subsection{Rotational diagrams}
\label{sec:rot-diagr}

\begin{figure*} 
  \centering 
  \includegraphics[width=17cm]{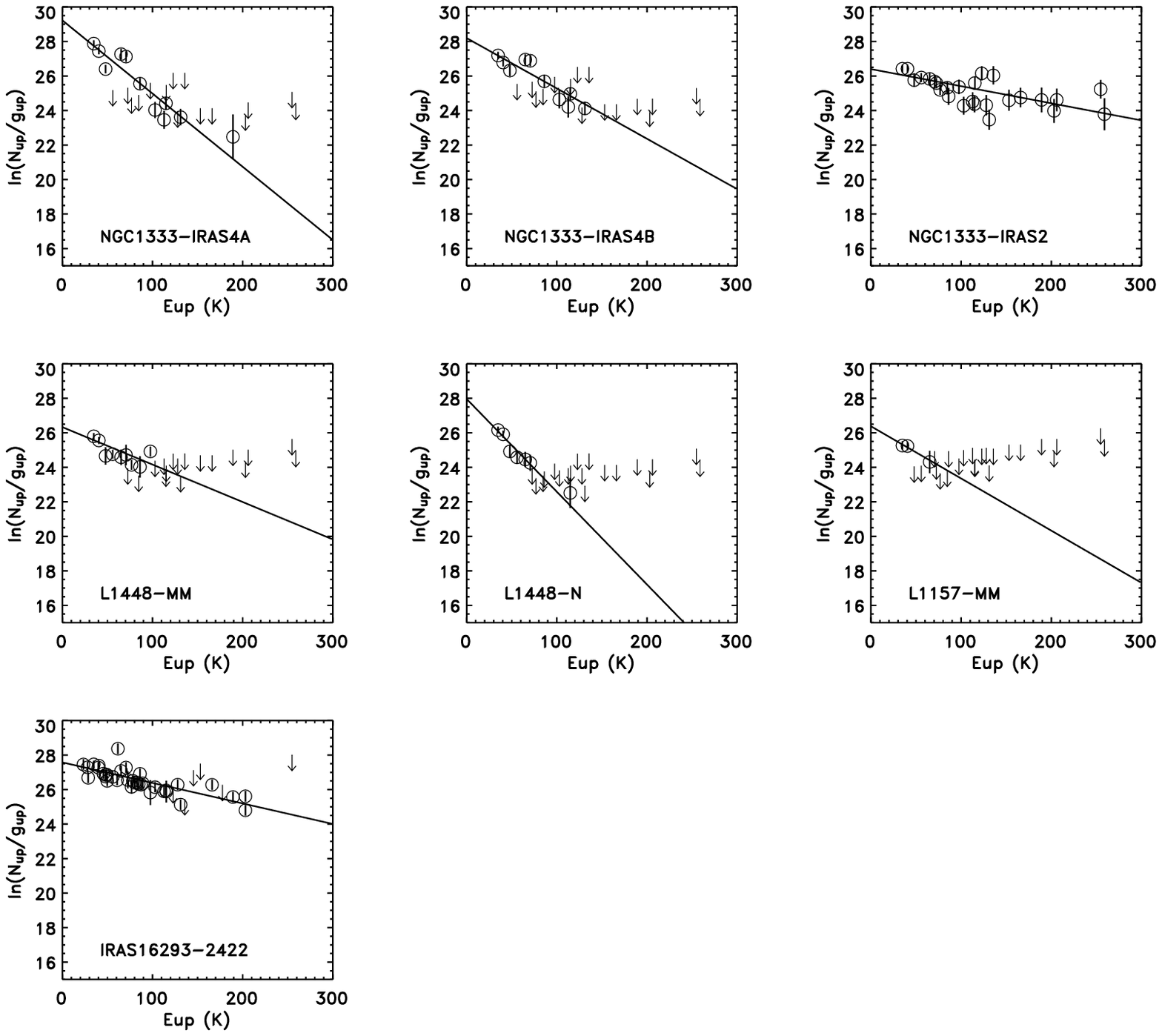} 
  \caption{Rotational diagrams of CH$_3$OH lines. Detected lines are
    marked by circles with error bars. Arrows indicate $2 \sigma$ upper
    limits.}
  \label{fig:rot-diag} 
\end{figure*}

Rough estimates of the kinetic temperatures and column densities have
been obtained from the rotational diagrams of each source
(Fig. \ref{fig:rot-diag}). The observations are reasonably well fitted
by a straight line, with some scattering, which is probably due to
opacity and non-LTE effects.  No differences are observed between the
JCMT and IRAM-30m observations, despite the different beam widths of
the two instruments.

The derived rotational temperature and column densities are summarized
in Table \ref{tab:rot-diag}.  The derived column densities range
between 0.5 and $8 \times 10^{14}\ \cmsqr$. For most of the sources,
rotational temperatures are around 30 K. Two notable exceptions are
NGC1333-IRAS2 and IRAS16293-2422, whose rotational temperatures are
about 100 and 85 K respectively.  That the temperatures are larger in
these two sources is not a surprise, for they are the only two sources
where several lines with high level energy are detected, despite the
fact that lines with low level energy are not much brighter than in
the other sources. It is therefore very likely that the observed
methanol lines in NGC1333-IRAS2 and IRAS16293-2422 originate in the
hot region of the envelope.

Rotational temperatures should, however, be considered with some
caution, because they may not reflect the actual kinetic
temperatures. As noted by \citet{Bachiller95} and \cite{Buckle02},
methanol can be very sub-thermally populated in interstellar
conditions, and the rotational temperature derived from methanol
rotational diagram analysis could be significantly lower than the
kinetic temperature. Part of the differences seen from one source to
an other could be due to differences in excitation. In the next
section, a more detailed analysis including non-LTE excitation is
developed, in order to take these effects into account.

\begin{table}
  \begin{center}
    \caption{Results from the rotational diagrams.}
    \label{tab:rot-diag}
    \begin{tabular}{l c c}          
      \hline
      \hline
      Source & $T_\mathrm{rot}$ & $N(\mathrm{CH}_3\mathrm{OH})$ \\
      & (K) & ($\cmsqr$) \\
      \hline
      NGC1333-IRAS4A & 24 $\pm$ 2 & $(5.1 \pm 1.0) \times 10^{14}$ \\
      NGC1333-IRAS4B & 34 $\pm$ 4 & $(3.5 \pm 0.8) \times 10^{14}$ \\
      NGC1333-IRAS2 & 101 $\pm$ 16 & $(3.4 \pm  0.6) \times 10^{14}$ \\
      L1448-MM & 46 $\pm$ 10 & $(8.8 \pm 2.4) \times 10^{13}$ \\
      L1448-N & 19 $\pm$ 3 & $(9.4 \pm 3.6) \times 10^{13}$ \\
      L1157-MM & 33 $\pm$ 25 & $(5.3 \pm  4.9) \times 10^{13}$ \\
      IRAS16293-2422 & 84 $\pm$ 8 & $(8.1 \pm 0.9) \times 10^{14}$ \\
      \hline
    \end{tabular}
  \end{center}
\end{table}

%%% Local Variables: 
%%% mode: latex
%%% TeX-master: "~/Methanol/ArticleAA/aa_ch3oh"
%%% End: 

\subsection{Jump models}
\label{sec:jump-models}

To determine more precisely the methanol abundance in the envelopes, a
detailed 1-D radiative transfer model has been developed. The model
uses the escape probability formalism, and has been presented in
detail in \citetalias{Maret04}. The energy levels and Einstein
coefficients were taken from the Cologne Molecular Database
Spectroscopy \citep{Muller01}. CH$_3$OH collisional rates with para
H$_2$ from \citet{Pottage04} were used. The threefold hindering
potential of the molecule leads to the formation of $A$ and $E$
symmetry states (see Appendix \ref{sec:spectral-designation}). Because
no radiative or collisional transitions are possible between the $E$
and $A$ symmetry species, these were modelled separately. The first
100 levels were considered for each symmetry of the molecule. Finally,
only the ground torsional vibration state ($v_t$=0) was
considered. The first excited state of the torsional vibration
($v_t$=1) is about 200 cm$^{-1}$ ($\sim$ 290 K) above the ground
state. Levels with $v_t$=1 might be excited and decay to the torsional
vibration ground state, but in a higher rotational state. However,
radiative excitation due to the dust is generally inefficient, and,
for the density and temperature ranges present here, collisional
excitation to those levels is expected to be negligible
\citep{Schoier02}.

The envelopes have been assumed to be spherically symmetric, and
power-law density profiles have been adopted. The index of the
power-law density profiles and the dust temperature profiles were
taken from \citet{Jorgensen02} and \citetalias{Maret04}. The methanol
abundance profile has been supposed to follow a step function
\citepalias{Maret04}. The abundance is set to a constant value
$\Xout$, in the outer cold part of the envelope. This abundance jumps
to a $\Xin$ value in the inner and hotter part of the envelope,
because of the evaporation of the grain mantles.

The adopted ``jump'' abundance profiles may seem simplistic, because
other physical and chemical processes in the envelope could lead to
more complex abundance profiles. It has been suggested that the
formaldehyde abundance in the outer envelope of L1448-C and
IRAS16293-2422 follow a ``drop'' profile \citep{Schoier04}. In this
model, the abundance is reduced in the cold dense part of the envelope
where CO is frozen out, but is undepleted in the inner region of the
envelope, where T $>$ 50 K. The outer radius of the ``drop'' region is
fixed where the density equals $10^5 \ \cmcub$. Support for this
suggestion stems from OVRO observations, which cannot be reproduced on
the short baselines by the jump scenario proposed in
\citetalias{Maret04}. Increasing the formaldehyde abundance in the
outer region was found to give a better agreement with the
observations.

Another possible explanation for the discrepancy between ``jump''
models and the OVRO observations could be that, for transitions with
the lowest level energy, the clouds in which the protostars are
embedded contribute significantly to the short baseline emission.
This emission was not taken into account either in
\citetalias{Maret04} or in \citet{Schoier04}. Emission from the cloud
would have the same effect as increasing the formaldehyde abundance in
the outer envelope. Indeed, the outer radii of the ``drop'' region is
about 5000 AU and 8 000 AU from the center for IRAS16293-2422 and
L1448-MM respectively. At these radius, the temperature of the gas is
only $\sim 10$ K, and the contribution of the cloud is likely to
become important.

Therefore, the main difference between the ``jump'' and ``drop''
scenarios is the temperature at which the evaporation occurs in the
inner region of the envelope. \citet{Schoier04} adopted a temperature
of 50 K, while a temperature of 90 K was adopted in
\citetalias{Maret04}. Methanol trapped in water ice will evaporate at
the sublimation temperature of pure water ice ($\sim 90$ K). Because
observations of interstellar ices suggests that CH$_3$OH is coexistent
with H$_2$O in the ices \citep{Skinner92,Gerakines99,Boogert00b}, we
have adopted 90 K for the sublimation temperature. We refer the reader
to \citetalias{Maret04} for a discussion of the effect of varying the
evaporation temperature.

In principle, the evaporation temperature depends on the composition
of ices. If methanol is trapped in water rich (polar) ices, the
evaporation is likely to occur at the evaporation temperature of pure
water ice ($\sim$ 90 K). On the other hand, if methanol is trapped in
CO rich (apolar) ices, the temperature will occur at a lower
temperature ($\sim$ 30 K). In this study, the methanol was assumed to
be trapped mostly in water rich ices, and the evaporation temperature
was therefore assumed to be 90 K. We refer the reader to
\citetalias{Maret04} for a discussion of the effect of varying the
evaporation temperature.

\begin{figure*} 
  \centering 
  \includegraphics[width=17cm]{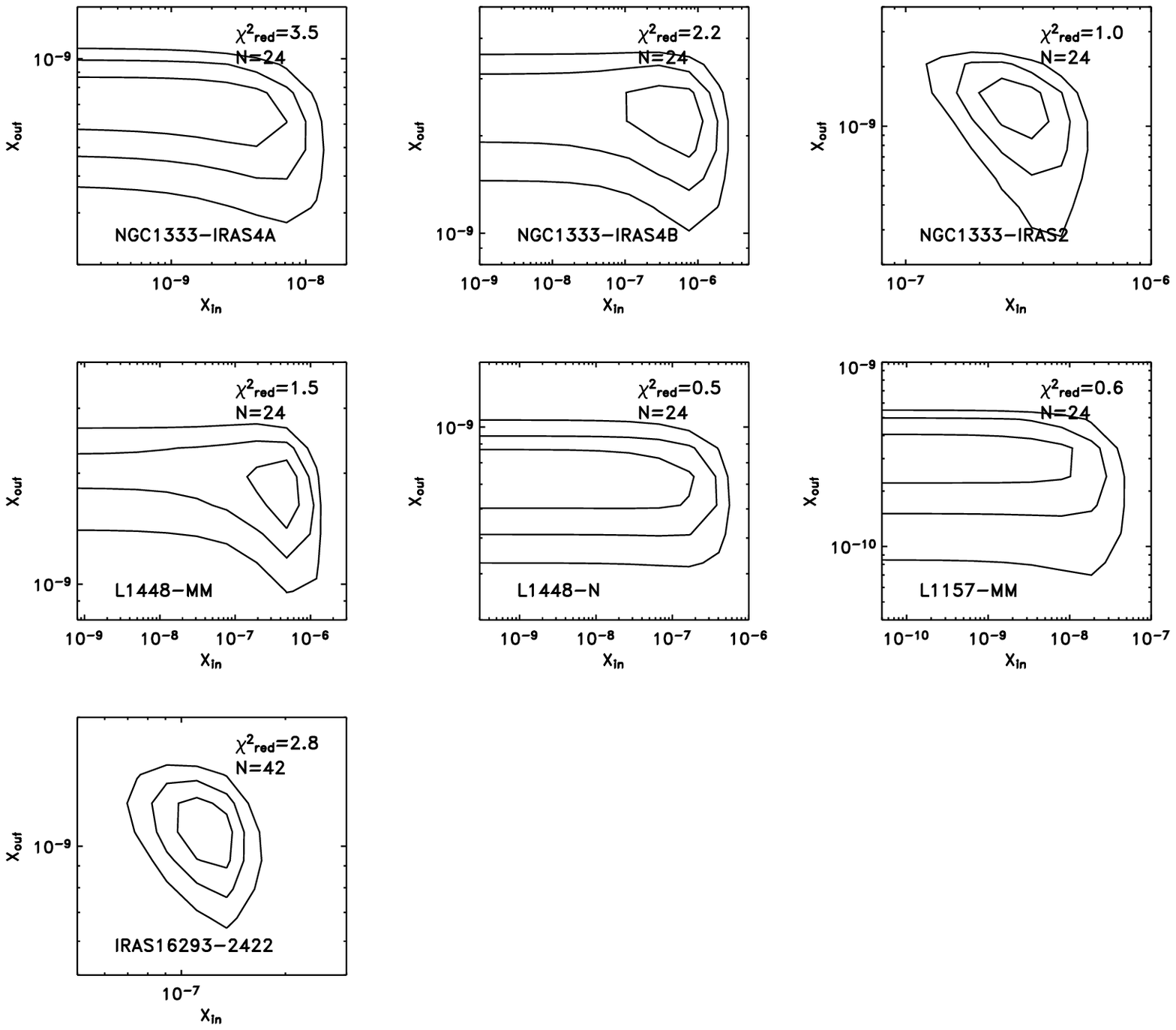} 
  \caption{$\chi^2$ as a function of the inner and outer CH$_3$OH
    abundance. The contours indicate the 1, 2 and 3$\sigma$ confidence
    levels. The number of observed lines, $N$, are also shown, as
    well as the reduced $\chi^2$ for the best fit model.
}
  \label{fig:chi2} 
\end{figure*}

The values of $\Xout$ and $\Xin$ were determined by running a grid of
10 $\times$ 10 models for each source, with the abundances ranging
from 10$^{-12}$ to 10$^{-4}$. The best fit parameters were then
determined by computing the $\chi^2$ at each point of the grid. The
grid was eventually refined around the best fit solutions to determine
more precisely the abundances. Fig. \ref{fig:chi2} present the
$\chi^2$ maps obtained for each source. The figure indicates the 1, 2
and 3$\sigma$ confidence levels, the number of lines considered, as
well as the reduced $\chi^2$ for the best fit model. The best fit
models reproduce the observations quite well, with the reduced
$\chi^2$ ranging from 0.5 to 3.5. The abundances obtained are
presented in Table \ref{tab:abundances}.

The methanol abundance in the outer part of the envelope ranges from
$3 \times 10^{-10}$ to $2 \times 10^{-9}$. The inner abundance varies
much more from one source to the other. In two sources of our sample,
NGC1333-IRAS2 and IRAS16293-2422, the observations can only be
reproduced by our model if there are jumps in the abundances.  The
inner abundances are $3 \times 10^{-7}$ and $1 \times 10^{-7}$
respectively, i.e. a factor 100 and 200 larger than in the outer
envelope. We note that IRAS16293-2422 abundances are in good agreement
with those obtained previously by \citet{Schoier02}. In NGC1333-IRAS4B
and L1448-MM, there is weak evidence for abundance jumps ($1
\sigma$). In these sources, the inner abundance jumps to $7 \times
10^{-7}$ and $5 \times 10^{-7}$ respectively, i.e. a factor about 300
larger than in the outer cold part of the envelope. These values
should however be treated with some caution, because of the low
confidence level. In NGC1333-IRAS4A, L1448-N and L1157-MM, the
observations are well reproduced with a constant CH$_3$OH abundance
throughout the envelope, even if the presence of a jump up to a few
$10^{-7}$ cannot be ruled out by the present observations in these
sources.

\begin{table*}
\begin{center}
  \caption{Summary of derived abundances.}
  \label{tab:abundances}
  \begin{tabular}{l c c c c c}
    \hline
    \hline

    Source & CO\tabnote{a} & \multicolumn{2}{c}{H$_2$CO\tabnote{b}} &
    \multicolumn{2}{c}{CH$_3$OH\tabnote{c}} \\
    
    && Outer\tabnote{d} & Inner\tabnote{e} & Outer\tabnote{d} &
    Inner\tabnote{e}\\
    
    \hline
    
    NGC1333-IRAS4A & $8 \times 10^{-6}$ & $2 \times 10^{-10}$ & $2
    \times 10^{-8}$ & $7 \times 10^{-10}$ & $< 1 \times 10^{-8}$ \\
    
    NGC1333-IRAS4B & $1 \times 10^{-5}$ & $5 \times 10^{-10}$ & $3
    \times 10^{-6}$ & $2 \times 10^{-9}$ & $7 \times
    10^{-7}$\tabnote{i}\\
    
    NGC1333-IRAS2 & $2 \times 10^{-5}$ & $3 \times 10^{-10}$ & $2 \times
    10^{-7}$ & $1 \times 10^{-9}$ & $3 \times 10^{-7}$ \\
    
    L1448-MM & $4 \times 10^{-5}$ & $7 \times 10^{-10}$ & $6 \times
    10^{-7}$ & $2 \times 10^{-9}$ & $5 \times
    10^{-7}$\tabnote{i}\\
    
    L1448-N & ... & $3 \times 10^{-10}$ & $1 \times 10^{-6}$ & $7
    \times 10^{-10}$ & $< 4 \times 10^{-7}$ \\
      
    L1157-MM & $6 \times 10^{-6}$ & $8 \times 10^{-11}$ & $1
    \times 10^{-8}$ & $3 \times 10^{-10}$ & $< 3 \times 10^{-8}$ \\
    
    IRAS16293-2422 & $3 \times 10^{-5}$ & $1 \times 10^{-9}$ & $1
    \times 10^{-7}$ & $1 \times 10^{-9}$ & $1 \times 10^{-7}$ \\
    
    \hline
    
    L134N\tabnote{d} & $1 \times 10^{-4}$ & \multicolumn{2}{c}{$2
      \times 10^{-8}$} & \multicolumn{2}{c}{$5 \times 10^{-9}$} \\
      
    Orion hot core\tabnote{e} & $1 \times 10^{-4}$ &
    \multicolumn{2}{c}{$7 \times 10^{-9}$} & \multicolumn{2}{c}{$1
      \times 10^{-7}$} \\

    High mass YSO\tabnote{f} & $(0.3 - 1.7) \times 10^{-4}$ &
    \multicolumn{2}{c}{$(1 - 10) \times 10^{-9}$} & $(0.4 - 24) \times
    10^{-9}$ & $(6 - 9) \times 10^{-8}$\\
  
    L1157 outflow\tabnote{g} & $1 \times 10^{-4}$ &
    \multicolumn{2}{c}{$2 \times 10^{-7}$} & \multicolumn{2}{c}{$2
      \times 10^{-5}$} \\
      
    Ices\tabnote{h} & $(0.2 - 3) \times 10^{-5}$ &
    \multicolumn{2}{c}{$(1 - 4) \times 10^{-6}$} &
    \multicolumn{2}{c}{$(0.2 - 2) \times 10^{-5}$} \\
      
%    Hale-Bopp comet\tabnote{k} & $1 \times 10^{-5}$ &
%    \multicolumn{2}{c}{$5 \times 10^{-7}$} & \multicolumn{2}{c}{$1
%      \times 10^{-6}$} \\
    
    \hline
    
  \end{tabular}
  
\end{center}

\tabnote{a} From \citet{Jorgensen04c} and \citet{Schoier02}.\\
\tabnote{b} From \citetalias{Maret04}.\\
\tabnote{c} This study.\\
\tabnote{d} From \citet{Dickens00} (position C).\\ % Not H2CO. Ohishi 1992 ?
\tabnote{e} From \citet{Sutton95}.\\
\tabnote{f} From \citet{vanderTak00b, vanderTak00a}.\\ % CO assumed to be 1e-4
\tabnote{g} From \citet{Bachiller97}.\\
\tabnote{h} From \citet{Ehrenfreund00} and \citet{Gibb04}, assuming
$X(\mathrm{H}_2\mathrm{O}) = 5 \times 10^{-5}$.\\
\tabnote{i} $1 \sigma$ confidence level only.

\end{table*}

%%% Local Variables: 
%%% mode: latex
%%% TeX-master: "/Users/smaret/Methanol/ArticleAA/aa_ch3oh"
%%% End: 

\section{Discussion}
\label{sec:discussion}

The abundance of methanol in the outer regions of Class 0 protostars
is $3-20 \times 10^{-10}$, a value comparable with the ones observed
in cold dark clouds \citep[$5 \times 10^{-9}$, ][]{Dickens00}. Also,
the observations reveal clear evidence for high methanol abundances in
the envelopes of two Class 0 protostars, IRAS16293-2422 and
NGC1333-IRAS2. At the two sigma level, two other protostars --
NGC1333-IRAS4A and L1448-MM -- also show enhanced CH$_3$OH abundances
in the inner regions. For the other sources, the evidence is
inconclusive.

The presence of large abundance jumps in the inner regions of low mass
protostars is often cited as evidence for the importance of
evaporation of ice mantles due to heating by the newly formed star.
These ice mantles are thought to form during the preceding, cold,
prestellar core phase. This scenario is supported by the detection of
high deuterium fractionation in the warm gas around low mass YSOs
\citep{Ceccarelli98b, Loinard02, Parise02, Parise04}, a tell-tale sign
of low temperature chemistry. The presence of complex organic
molecules (\eg CH$_3$OCH$_3$, HCOOCH$_3$) also shows the importance of
ice evaporation, as these molecules are generally thought to form from
abundant ice species through rapid reactions in the warm gas
\citep{Charnley92, Caselli93}. Methanol plays a central role in this
drive towards molecular complexity.

It is clear that the abundance of methanol in the warm inner regions
of low mass protostars can provide important clues to the chemistry of
regions of star formation and, in particular, to the importance of
grain surface chemistry versus gas phase chemistry.  In the remainder
of this section, we discuss gas phase and grain surface formation
routes of methanol and compare the abundances derived for the hot core
regions with models predictions.

In the gas phase, the main methanol formation route is through
radiative association of CH$_3^+$ with H$_2$O, followed by
dissociative recombination of CH$_3$OH$_2^+$ (Herbst, private
communication):

\begin{eqnarray}
  \mathrm{CH}_3^+ + \mathrm{H}_2\mathrm{O} \rightarrow
  \mathrm{CH}_3\mathrm{OH}_2^+ + h\nu \\
  \mathrm{CH}_3\mathrm{OH}_2^+ + e^- \rightarrow
  \mathrm{CH}_3\mathrm{OH} + \mathrm{H}
\end{eqnarray}

The bottleneck in this sequence is the first step, the radiative
association reaction.  The rate of this reaction was recently measured
to be three orders of magnitude lower at 10 K than previous
estimates. Using the earlier estimate, \citet{Roberts04} predicted a
methanol abundance in dense interstellar cores of 10$^{-8}$ at early
times (10$^{5}$ yr), decreasing to 10$^{-9}$ at later times (10$^{7}$
yr). Decreasing the reaction rate by about three orders of magnitude
is expected to decrease the methanol abundance by the same factor,
i.e. to between 10$^{-12}$ and 10$^{-11}$.

On grain surfaces, methanol is thought to be formed by successive
hydrogenation of CO \citep{Tielens82, Tielens87}:

\begin{equation}
  \label{eq:co_hydrogen}
  \mathrm{CO} \rightarrow \mathrm{HCO} \rightarrow
  \mathrm{H}_{2}\mathrm{CO} \rightarrow \mathrm{H}_{2}\mathrm{COH}
  \rightarrow \mathrm{CH}_{3}\mathrm{OH}
\end{equation}

This process has been studied experimentally by \citet{Hidaka04}, and
was found to be efficient forming methanol at low temperature.
Theoretical models show that methanol is abundant in icy grain mantles
(CH$_3$OH/H$_2$O$\sim 0.1$) when the atomic hydrogen abundance in the
accreting gas is high compared to the CO abundance \citep{Keane05}.

For IRAS16293-2422 and NGC1333-IRAS2, the methanol abundance in the
hot core compares well with those found in hot cores associated with
regions of massive star formation \citep[$6 - 10 \times 10^{-8}$,
][]{Sutton95, vanderTak00b, vanderTak00a} and are much higher than
methanol abundances observed in cold dark clouds \citep[$5 \times
10^{-9}$, ][]{Dickens00}. In contrast, much higher methanol abundances
are associated with the outflows in the regions of low mass star
formation, L1157-MM and NGC1333-IRAS2 \citep[$2 \times 10^{-5}$ and $2
\times 10^{-6}$ respectively, ][]{Bachiller97, Bachiller98}. Likewise,
cold ices observed along the line of sight towards high mass and low
mass protostars have methanol abundances ($0.2 - 2 \times 10^{-5}$)
which are two orders of magnitude larger than observed in hot cores
\citep{Allamandola92, Gibb04, Pontoppidan03, Pontoppidan04}.  Lines of
sight towards background stars on the other hand have much lower
methanol ice abundances (\citealp[$ < 5 \times 10^{-6}$,
][]{Chiar96}).

It was realized more than a decade ago that the methanol abundances
observed in the hot core associated with low and high mass protostars,
interstellar ices and YSO outflows pose grave challenges to gas phase
models for the formation of methanol \citep[\eg][]{Menten88}. Adopting
the new rate for the radiative association rate, gas phase models fall
short by four orders of magnitude.  Indeed, this is one (additional)
reason why the high methanol abundances in the warm gas are generally
thought to reflect release into the gas phase of grain mantle species
in the inner region of the envelope, either due to thermal evaporation
as a result of the heating by the protostar, or due to grain mantle
sputtering by the outflow. As discussed previously, our maps suggest
that the methanol lines with the highest level energy do not spatially
correspond to the outflow emission, but are rather peaked on the
central source. Together with the small line width, this suggests that
this emission originates in the hot core where grain mantles evaporate
completely when the dust temperature exceeds 90 K. A similar
conclusion was reached for the high H$_2$CO abundances observed in
these same sources \citepalias{Maret04}. Recently, \citet{Schoier04}
investigated the origin of H$_2$CO abundance enhancement, using
interferometric observations of IRAS16293-2422 and L1448-MM. Their
maps reveals a compact emission region and small line widths and are
not associated with the outflows. They conclude that these
enhancements originate from thermal evaporation of the grain mantles,
as claimed in \citetalias{Maret04}. Further support for the thermal
evaporation thesis is provided by the interferometric images of
complex organic molecules in IRAS16293-2422, which show compact
emission around the two protostars of the binary system, on scales
smaller than 1" \citep{Kuan04,Bottinelli04b}. Here, we conclude that
the scenario is very likely the same for both methanol and
formaldehyde.

If the abundances observed in the hot cores of protostars are due to
grain mantle evaporation, they should reflect the grain mantle
composition. The observed discrepancy in the ice ($0.2 - 2 \times
10^{-5}$) and hot core methanol abundances ($1 - 7 \times 10^{-7}$) is
then somewhat discomforting. Possibly, ice mantles are only partially
released into the gas phase in these regions or the gas phase of hot
cores is more heavily processed by gas phase reactions than realized
before \citep{Charnley92}. Such a partial release of ice mantles may
result from small differences in temperatures between different size
grains (\eg, smaller grains or graphitic grains might be slightly
warmer than larger silicate grains).  Typically, a 10\%\ difference in
temperature can cause a factor 5 difference in the evaporation rate of
ice mantles under interstellar conditions. Of course, within this
scenario, this would imply that these warmer grains represent only a
small fraction of the total ice reservoir. For shocks, partial release
of the ice mantles may be a natural outcome of the sputtering in low
velocity shocks.

The grain mantle scenario can be further tested by considering
observed abundance ratios, involving CH$_3$OH, H$_2$CO and CO rather
than absolute abundances. Of course, such a comparison inherently
assumes that these species evaporate simultaneously.  For traces of
CO, CH$_3$OH and H$_2$CO mixed into a H$_2$O-rich ice, evaporation
will be domainted by the evaporation of the main component. However,
some of the solid CO may not be coexisting with the H$_2$O-rich ice
mantle \citep{Tielens91,Chiar98,Pontoppidan03} and this rather
volatile ice component will evaporate more readily
\citep{Sandford93}. These uncertainties have to be kept in mind in the
subsequent discussion.

\begin{figure}
  \centering 
    \includegraphics[width=\columnwidth]{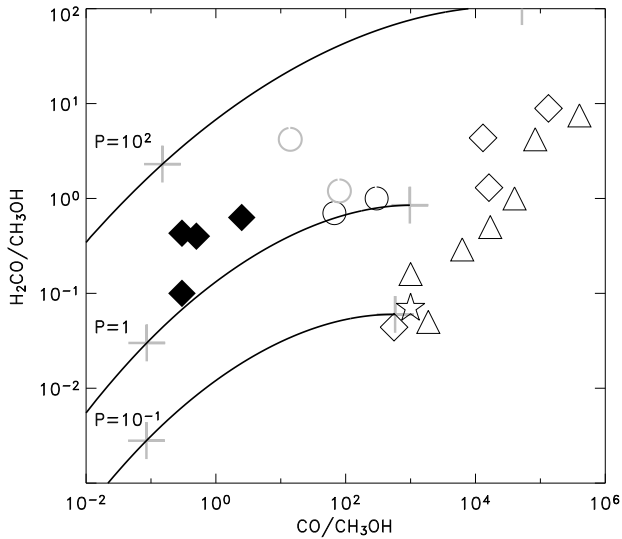} 
    \caption{Abundances ratio of H$_2$CO and CO relative to CH$_3$OH.
      The curves show the grain chemistry model predictions for
      different ratio of the probability $P$ of H to react with CO
      relative to H$_2$CO. Along each curve, the density
      increases. The crosses denote densities of 10$^4$ and 10$^5$
      cm$^{-3}$. Black diamonds indicates observed ice abundances
      towards a number of embedded young stellar objects
      \citep{Keane05}. Open diamonds represent the gas phase
      abundances observed in the same objects.  The star denotes the
      gas phase composition of the Orion hot core
      \citep{Sutton95}. The open circles indicate gas phase
      observations of low mass protostars (this study). Black circles
      represent IRAS16293-2422 and NGC1333-IRAS2, for which the jumps
      detections are firm, while grey circles represent NGC1333-IRAS4B
      and L1448-MM. The triangles indicate the observed gas phase
      abundances towards high mass YSO
      \citep{vanderTak00b,vanderTak00a}. Adapted from
      \citet{Keane05}.}
  \label{fig:grain_chem} 
\end{figure}

The formation of formaldehyde and methanol from CO and H accreted onto
grain surfaces has recently been studied theoretically by
\citet{Keane05}. In their model, the formaldehyde and methanol
abundances produced depend on two parameters: the density and the
ratio $P$ of the probability of H to react with CO relative to H$_2$CO
(see Eq. \ref{eq:co_hydrogen}). Fig. \ref{fig:grain_chem} presents the
abundances ratio of H$_2$CO and CO relative to CH$_3$OH predicted by
the model and observed in various environments.  First, we note that
the H$_2$CO/CH$_3$OH and CO/CH$_3$OH abundance ratios observed in the
warm gas around low mass protostars are somewhat higher than those
observed in ices (Fig. \ref{fig:grain_chem}). Both sets of
observations are well fitted by the model, for an adopted $P$ of 1.
The differences between the low mass hot cores and the high mass ices
reflect a difference in the density during accretion, 10$^{5}$ versus
$10^4$ cm$^{-3}$, respectively.  This value for $P$ is in good
agreement with recent experimental estimates \citep{Hidaka04}. We note
that for a density of $10^5$ cm$^{-3}$, the calculated CH$_3$OH/H$_2$O
ratio is $10^{-3}$. This is about a factor of 40 less than the
observed CH$_3$OH/H$_2$O ratio in IRAS16293-2422 (assuming a water
abundance of $3 \times 10^{-6}$, \citealt{Ceccarelli00a}). Higher
abundance ratios are obtained when accretion occurs at lower densities
where the atomic H abundance is higher \citep{Keane05}.  Indeed, both
the observed absolute and relative abundances of these species are in
good agreement with the ice observations towards high mass protostars
for a density of 10$^{4}$ cm$^{-3}$ at the time of formation
\citep{Keane05}.

We can also compare the observed abundance ratios in low mass hot
cores with those observed in high mass hot cores
(Fig. \ref{fig:grain_chem}).  This comparison reveals that relative to
CH$_3$OH, the CO abundance is typically enhanced by an order of
magnitude while the H$_2$CO abundance is typically less by a factor 3
in the latter compared to the former. As already noted by
\citet{Keane05}, the hot core composition in regions of massive star
formation are quite different from those observed in the ices towards
the same type of objects (\cf Fig. \ref{fig:grain_chem}). Possibly,
the composition of the hot core has already been substantially altered
from that of the evaporating ices by reactions in the warm gas
\citep{Keane05}. Hot core chemistry models predict that the abundance
of H$_2$CO decreases before CH$_3$OH \citep{Charnley92}, so the
H$_2$CO over CH$_3$OH ratio is expected to decrease with time. Hence,
the points in Fig \ref{fig:grain_chem} are expected to shift from the
calculated grain surface chemistry curves to the lower right corner
with time.  From this perspective, the differences in the abundance
ratios between high and low mass hot cores could therefore be
interpreted in terms of different ages for these two kind of regions:
high mass hot cores are typically older than low mass hot core. Gas
phase chemistry calculations are needed to confirm this scenario.

Summarizing this discussion, we conclude that current gas phase
chemistry models fail to reproduce the observed methanol abundances by
many orders of magnitude.  Based upon recent experimental and
theoretical studies, good agreement is obtained between the
observations and grain surface chemistry models in terms of relative
abundance ratios of CH$_3$OH, H$_2$CO and CO. However, reproduction of
absolute abundances seems to be more troublesome, perhaps pointing
towards the presence of multiple ice components and for selective
partial evaporation. While this comparison presents, at the moment,
mixed results, further observational studies of these and related
species, in combination with laboratory studies and theoretical
models, promises the opportunity of developing a chemical clock which
can ``time'' the star formation process.

\section{Conclusions}
\label{sec:conclusion}

Methanol lines observations of a sample of Class 0 protostars have
been presented. Using a detailed 1-D radiative transfer model, the
methanol abundance profiles have been derived and compared with
chemical scenarios of the formation of this molecule. Our main
conclusions are:

\begin{enumerate}
  
  \item The methanol abundances in the outer cold envelopes ($<$ 90 K)
    are 3-20 $\times$ 10$^{-10}$, a comparable value with the ones
    observed in cold dark clouds.

  \item Constant abundance models fail to reproduce the observations
    in two sources of our sample, IRAS16293-2422 and NGC1333-IRAS2. A
    jump at the radius where the temperature reaches 90 K has to be
    included in order to reproduce the observations. In two other
    sources, NGC1333-IRAS4B and L1448-MM, jumps in the abundance are
    also detected, but at a lower confidence level. In the three
    remaining sources, NGC1333-IRAS4A, L1448-N and L1157-MM, the
    observations are well reproduced by a constant abundance model,
    but the presence of a jump in the abundance cannot be ruled out.

  \item In the sources where abundances jumps are detected, the
    abundance in the inner hot region ($>$ 90 K) are 1-7 $\times$
    10$^{-7}$, i.e. about two orders of magnitude larger than in cold
    dark cloud. The abundances in these hot cores compare well with
    those derived for the Orion hot core, as well as in the inner
    region of high mass YSO.

  \item We have compared the observed methanol abundances with gas
    phase and grain surface chemistry models. Gas phase chemistry
    models predict methanol abundances (10$^{-11}$) much lower than
    observed in these hot cores. Grain surface chemistry models
    provide good agreement with the observed abundance ratios
    involving CH$_3$OH, H$_2$CO, and CO.  However, absolute abundances
    are more difficult to reproduce and may point towards the presence
    of multiple ice components and the effects of difference in
    volatility between different ices species.
    
\end{enumerate}

\acknowledgements{The authors are grateful to Ewine van Dishoeck, Jes
  J\o rgensen and Fredrik Sch\"oier for useful discussions about the
  formaldehyde and methanol abundances. Eric Herbst and Ted Bergin are
  also thanked for discussion about the methanol formation. The
  authors which also to thanks the JCMT and IRAM-30m telescope staff,
  for their technical help during the observations presented in the
  paper. We thanks the referee for useful comments on this manuscript.
}

\bibliography{bibliography} 

\begin{thebibliography}{60}
\expandafter\ifx\csname natexlab\endcsname\relax\def\natexlab#1{#1}\fi

\bibitem[{{Allamandola} {et~al.}(1992){Allamandola}, {Sandford}, {Tielens}, \&
  {Herbst}}]{Allamandola92}
{Allamandola}, L.~J., {Sandford}, S.~A., {Tielens}, A.~G.~G.~M., \& {Herbst},
  T.~M. 1992, \apj, 399, 134

\bibitem[{{Andr{\'e}} {et~al.}(2000){Andr{\'e}}, {Ward-Thompson}, \&
  {Barsony}}]{Andre00}
{Andr{\'e}}, P., {Ward-Thompson}, D., \& {Barsony}, M. 2000, Protostars and
  Planets IV, 59+

\bibitem[{{Bachiller} {et~al.}(1998){Bachiller}, {Codella}, {Colomer},
  {Liechti}, \& {Walmsley}}]{Bachiller98}
{Bachiller}, R., {Codella}, C., {Colomer}, F., {Liechti}, S., \& {Walmsley},
  C.~M. 1998, \aap, 335, 266

\bibitem[{{Bachiller} {et~al.}(1995){Bachiller}, {Guilloteau}, {Dutrey},
  {Planesas}, \& {Martin-Pintado}}]{Bachiller95}
{Bachiller}, R., {Guilloteau}, S., {Dutrey}, A., {Planesas}, P., \&
  {Martin-Pintado}, J. 1995, \aap, 299, 857

\bibitem[{{Bachiller} \& {Perez Gutierrez}(1997)}]{Bachiller97}
{Bachiller}, R. \& {Perez Gutierrez}, M. 1997, \apjl, 487, L93+

\bibitem[{{Blake}(1996)}]{Blake96}
{Blake}, G.~A. 1996, in IAU Symp. 178: Molecules in Astrophysics: Probes \&
  Processes, 31--+

\bibitem[{{Blake} {et~al.}(1995){Blake}, {Sandell}, {van Dishoeck},
  {Groesbeck}, {Mundy}, \& {Aspin}}]{Blake95}
{Blake}, G.~A., {Sandell}, G., {van Dishoeck}, E.~F., {et~al.} 1995, \apj, 441,
  689

\bibitem[{{Boogert} {et~al.}(2000){Boogert}, {Tielens}, {Ceccarelli},
  {Boonman}, {van Dishoeck}, {Keane}, {Whittet}, \& {de Graauw}}]{Boogert00b}
{Boogert}, A.~C.~A., {Tielens}, A.~G.~G.~M., {Ceccarelli}, C., {et~al.} 2000,
  \aap, 360, 683

\bibitem[{{Bottinelli} {et~al.}(2004{\natexlab{a}}){Bottinelli}, {Ceccarelli},
  {Lefloch}, {Williams}, {Castets}, {Caux}, {Cazaux}, {Maret}, {Parise}, \&
  {Tielens}}]{Bottinelli04a}
{Bottinelli}, S., {Ceccarelli}, C., {Lefloch}, B., {et~al.} 2004{\natexlab{a}},
  \apj, 615, 354

\bibitem[{{Bottinelli} {et~al.}(2004{\natexlab{b}}){Bottinelli}, {Ceccarelli},
  {Neri}, {Williams}, {Caux}, {Cazaux}, {Lefloch}, {Maret}, \&
  {Tielens}}]{Bottinelli04b}
{Bottinelli}, S., {Ceccarelli}, C., {Neri}, R., {et~al.} 2004{\natexlab{b}},
  \apjl, 617, L69

\bibitem[{{Buckle} \& {Fuller}(2002)}]{Buckle02}
{Buckle}, J.~V. \& {Fuller}, G.~A. 2002, \aap, 381, 77

\bibitem[{{Caselli} {et~al.}(1993){Caselli}, {Hasegawa}, \&
  {Herbst}}]{Caselli93}
{Caselli}, P., {Hasegawa}, T.~I., \& {Herbst}, E. 1993, \apj, 408, 548

\bibitem[{{Cazaux} {et~al.}(2003){Cazaux}, {Tielens}, {Ceccarelli}, {Castets},
  {Wakelam}, {Caux}, {Parise}, \& {Teyssier}}]{Cazaux03}
{Cazaux}, S., {Tielens}, A.~G.~G.~M., {Ceccarelli}, C., {et~al.} 2003, \apjl,
  539, L51

\bibitem[{{Ceccarelli} {et~al.}(2000{\natexlab{a}}){Ceccarelli}, {Castets},
  {Caux}, {Hollenbach}, {Loinard}, {Molinari}, \& {Tielens}}]{Ceccarelli00a}
{Ceccarelli}, C., {Castets}, A., {Caux}, E., {et~al.} 2000{\natexlab{a}}, \aap,
  355, 1129

\bibitem[{{Ceccarelli} {et~al.}(1998){Ceccarelli}, {Castets}, {Loinard},
  {Caux}, \& {Tielens}}]{Ceccarelli98b}
{Ceccarelli}, C., {Castets}, A., {Loinard}, L., {Caux}, E., \& {Tielens},
  A.~G.~G.~M. 1998, \aap, 338, L43

\bibitem[{{Ceccarelli} {et~al.}(2000{\natexlab{b}}){Ceccarelli}, {Loinard},
  {Castets}, {Tielens}, \& {Caux}}]{Ceccarelli00b}
{Ceccarelli}, C., {Loinard}, L., {Castets}, A., {Tielens}, A.~G.~G.~M., \&
  {Caux}, E. 2000{\natexlab{b}}, \aap, 357, L9

\bibitem[{{\v{C}ernis}(1990)}]{Cernis90}
{\v{C}ernis}, K. 1990, \apss, 166, 315

\bibitem[{{Charnley} {et~al.}(1992){Charnley}, {Tielens}, \&
  {Millar}}]{Charnley92}
{Charnley}, S.~B., {Tielens}, A.~G.~G.~M., \& {Millar}, T.~J. 1992, \apjl, 399,
  L71

\bibitem[{{Chiar} {et~al.}(1996){Chiar}, {Adamson}, \& {Whittet}}]{Chiar96}
{Chiar}, J.~E., {Adamson}, A.~J., \& {Whittet}, D.~C.~B. 1996, \apj, 472, 665

\bibitem[{{Chiar} {et~al.}(1998){Chiar}, {Gerakines}, {Whittet}, {Pendleton},
  {Tielens}, {Adamson}, \& {Boogert}}]{Chiar98}
{Chiar}, J.~E., {Gerakines}, P.~A., {Whittet}, D.~C.~B., {et~al.} 1998, \apj,
  498, 716

\bibitem[{{Dickens} {et~al.}(2000){Dickens}, {Irvine}, {Snell}, {Bergin},
  {Schloerb}, {Pratap}, \& {Miralles}}]{Dickens00}
{Dickens}, J.~E., {Irvine}, W.~M., {Snell}, R.~L., {et~al.} 2000, \apj, 542,
  870

\bibitem[{{Ehrenfreund} \& {Charnley}(2000)}]{Ehrenfreund00}
{Ehrenfreund}, P. \& {Charnley}, S.~B. 2000, \araa, 38, 427

\bibitem[{{Gerakines} {et~al.}(1999){Gerakines}, {Whittet}, {Ehrenfreund},
  {Boogert}, {Tielens}, {Schutte}, {Chiar}, {van Dishoeck}, {Prusti},
  {Helmich}, \& {de Graauw}}]{Gerakines99}
{Gerakines}, P.~A., {Whittet}, D.~C.~B., {Ehrenfreund}, P., {et~al.} 1999,
  \apj, 522, 357

\bibitem[{{Gibb} {et~al.}(2004){Gibb}, {Whittet}, {Boogert}, \&
  {Tielens}}]{Gibb04}
{Gibb}, E.~L., {Whittet}, D.~C.~B., {Boogert}, A.~C.~A., \& {Tielens},
  A.~G.~G.~M. 2004, \apjs, 151, 35

\bibitem[{{Hidaka} {et~al.}(2004){Hidaka}, {Watanabe}, {Shiraki}, {Nagaoka}, \&
  {Kouchi}}]{Hidaka04}
{Hidaka}, H., {Watanabe}, N., {Shiraki}, T., {Nagaoka}, A., \& {Kouchi}, A.
  2004, \apj, 614, 1124

\bibitem[{{J{\o}rgensen} {et~al.}(2004{\natexlab{a}}){J{\o}rgensen},
  {Hogerheijde}, {Blake}, {van Dishoeck}, {Mundy}, \& {Sch{\"
  o}ier}}]{Jorgensen04b}
{J{\o}rgensen}, J.~K., {Hogerheijde}, M.~R., {Blake}, G.~A., {et~al.}
  2004{\natexlab{a}}, \aap, 415, 1021

\bibitem[{{J{\o}rgensen} {et~al.}(2002){J{\o}rgensen}, {Sch{\" o}ier}, \& {van
  Dishoeck}}]{Jorgensen02}
{J{\o}rgensen}, J.~K., {Sch{\" o}ier}, F.~L., \& {van Dishoeck}, E.~F. 2002,
  \aap, 389, 908

\bibitem[{{J{\o}rgensen} {et~al.}(2004{\natexlab{b}}){J{\o}rgensen}, {Sch{\"
  o}ier}, \& {van Dishoeck}}]{Jorgensen04c}
{J{\o}rgensen}, J.~K., {Sch{\" o}ier}, F.~L., \& {van Dishoeck}, E.~F.
  2004{\natexlab{b}}, \aap, 416, 603

\bibitem[{Keane \& Tielens(2005)}]{Keane05}
Keane, J.~V. \& Tielens, A.~G.~G. 2005, \aap, accepted

\bibitem[{{Knee} \& {Sandell}(2000)}]{Knee00}
{Knee}, L.~B.~G. \& {Sandell}, G. 2000, \aap, 361, 671

\bibitem[{{Kuan} {et~al.}(2004){Kuan}, {Huang}, {Charnley}, {Hirano},
  {Takakuwa}, {Wilner}, {Liu}, {Ohashi}, {Bourke}, {Qi}, \& {Zhang}}]{Kuan04}
{Kuan}, Y., {Huang}, H., {Charnley}, S.~B., {et~al.} 2004, \apjl, 616, L27

\bibitem[{{Langer} {et~al.}(1996){Langer}, {Castets}, \& {Lefloch}}]{Langer96}
{Langer}, W.~D., {Castets}, A., \& {Lefloch}, B. 1996, \apjl, 471, L111

\bibitem[{{Lees}(1973)}]{Lees73}
{Lees}, R.~M. 1973, \apj, 184, 763

\bibitem[{Lees \& Baker(1968)}]{Lees68}
Lees, R.~M. \& Baker, J. 1968, J. Chem. Phys, 48, 5299

\bibitem[{{Lefloch} {et~al.}(1998){Lefloch}, {Castets}, {Cernicharo}, {Langer},
  \& {Zylka}}]{Lefloch98}
{Lefloch}, B., {Castets}, A., {Cernicharo}, J., {Langer}, W.~D., \& {Zylka}, R.
  1998, \aap, 334, 269

\bibitem[{{Leurini} {et~al.}(2004){Leurini}, {Schilke}, {Menten}, {Flower},
  {Pottage}, \& {Xu}}]{Leurini04}
{Leurini}, S., {Schilke}, P., {Menten}, K.~M., {et~al.} 2004, \aap, 422, 573

\bibitem[{{Loinard} {et~al.}(2002){Loinard}, {Castets}, {Ceccarelli},
  {Lefloch}, {Benayoun}, {Caux}, {Vastel}, {Dartois}, \& {Tielens}}]{Loinard02}
{Loinard}, L., {Castets}, A., {Ceccarelli}, C., {et~al.} 2002, \planss, 50,
  1205

\bibitem[{{M{\" u}ller} {et~al.}(2001){M{\" u}ller}, {Thorwirth}, {Roth}, \&
  {Winnewisser}}]{Muller01}
{M{\" u}ller}, H.~S.~P., {Thorwirth}, S., {Roth}, D.~A., \& {Winnewisser}, G.
  2001, \aap, 370, L49

\bibitem[{{Maret} {et~al.}(2002){Maret}, {Ceccarelli}, {Caux}, {Tielens}, \&
  {Castets}}]{Maret02}
{Maret}, S., {Ceccarelli}, C., {Caux}, E., {Tielens}, A.~G.~G.~M., \&
  {Castets}, A. 2002, \aap, 395, 573

\bibitem[{{Maret} {et~al.}(2004){Maret}, {Ceccarelli}, {Caux}, {Tielens}, {J{\"
  o}rgensen}, {van Dishoeck}, {Bacmann}, {Castets}, {Lefloch}, {Loinard},
  {Parise}, \& {Sch{\" o}ier}}]{Maret04}
{Maret}, S., {Ceccarelli}, C., {Caux}, E., {et~al.} 2004, \aap, 416, 577

\bibitem[{{Menten} {et~al.}(1988){Menten}, {Walmsley}, {Henkel}, \&
  {Wilson}}]{Menten88}
{Menten}, K.~M., {Walmsley}, C.~M., {Henkel}, C., \& {Wilson}, T.~L. 1988,
  \aap, 198, 253

\bibitem[{{Menten} {et~al.}(1986){Menten}, {Walmsley}, {Henkel}, {Wilson},
  {Snyder}, {Hollis}, \& {Lovas}}]{Menten86}
{Menten}, K.~M., {Walmsley}, C.~M., {Henkel}, C., {et~al.} 1986, \aap, 169, 271

\bibitem[{{Parise} {et~al.}(2004){Parise}, {Castets}, {Herbst}, {Caux},
  {Ceccarelli}, {Mukhopadhyay}, \& {Tielens}}]{Parise04}
{Parise}, B., {Castets}, A., {Herbst}, E., {et~al.} 2004, \aap, 416, 159

\bibitem[{{Parise} {et~al.}(2002){Parise}, {Ceccarelli}, {Tielens}, {Herbst},
  {Lefloch}, {Caux}, {Castets}, {Mukhopadhyay}, {Pagani}, \&
  {Loinard}}]{Parise02}
{Parise}, B., {Ceccarelli}, C., {Tielens}, A.~G.~G.~M., {et~al.} 2002, \aap,
  393, L49

\bibitem[{{Pontoppidan} {et~al.}(2003){Pontoppidan}, {Fraser}, {Dartois},
  {Thi}, {van Dishoeck}, {Boogert}, {d'Hendecourt}, {Tielens}, \&
  {Bisschop}}]{Pontoppidan03}
{Pontoppidan}, K.~M., {Fraser}, H.~J., {Dartois}, E., {et~al.} 2003, \aap, 408,
  981

\bibitem[{{Pontoppidan} {et~al.}(2004){Pontoppidan}, {van Dishoeck}, \&
  {Dartois}}]{Pontoppidan04}
{Pontoppidan}, K.~M., {van Dishoeck}, E.~F., \& {Dartois}, E. 2004, \aap, 426,
  925

\bibitem[{{Pottage} {et~al.}(2004){Pottage}, {Flower}, \& {Davis}}]{Pottage04}
{Pottage}, J.~T., {Flower}, D.~R., \& {Davis}, S.~L. 2004, \mnras, 352, 39

\bibitem[{{Roberts} {et~al.}(2004){Roberts}, {Herbst}, \& {Millar}}]{Roberts04}
{Roberts}, H., {Herbst}, E., \& {Millar}, T.~J. 2004, \aap, 424, 905

\bibitem[{{Rodgers} \& {Charnley}(2003)}]{Rodgers03}
{Rodgers}, S.~D. \& {Charnley}, S.~B. 2003, \apj, 585, 355

\bibitem[{{Sandford} \& {Allamandola}(1993)}]{Sandford93}
{Sandford}, S.~A. \& {Allamandola}, L.~J. 1993, \apj, 417, 815

\bibitem[{{Sch{\" o}ier} {et~al.}(2002){Sch{\" o}ier}, {J{\o}rgensen}, {van
  Dishoeck}, \& {Blake}}]{Schoier02}
{Sch{\" o}ier}, F.~L., {J{\o}rgensen}, J.~K., {van Dishoeck}, E.~F., \&
  {Blake}, G.~A. 2002, \aap, 390, 1001

\bibitem[{{Sch{\" o}ier} {et~al.}(2004){Sch{\" o}ier}, {J{\o}rgensen}, {van
  Dishoeck}, \& {Blake}}]{Schoier04}
{Sch{\" o}ier}, F.~L., {J{\o}rgensen}, J.~K., {van Dishoeck}, E.~F., \&
  {Blake}, G.~A. 2004, \aap, 418, 185

\bibitem[{{Skinner} {et~al.}(1992){Skinner}, {Tielens}, {Barlow}, \&
  {Justtanont}}]{Skinner92}
{Skinner}, C.~J., {Tielens}, A.~G.~G.~M., {Barlow}, M.~J., \& {Justtanont}, K.
  1992, \apjl, 399, L79

\bibitem[{{Sutton} {et~al.}(1995){Sutton}, {Peng}, {Danchi}, {Jaminet},
  {Sandell}, \& {Russell}}]{Sutton95}
{Sutton}, E.~C., {Peng}, R., {Danchi}, W.~C., {et~al.} 1995, \apjs, 97, 455

\bibitem[{{Tielens} \& {Allamandola}(1987)}]{Tielens87}
{Tielens}, A.~G.~G.~M. \& {Allamandola}, L.~J. 1987, in NATO ASIC Proc. 210:
  Physical Processes in Interstellar Clouds, 333--376

\bibitem[{{Tielens} \& {Hagen}(1982)}]{Tielens82}
{Tielens}, A.~G.~G.~M. \& {Hagen}, W. 1982, \aap, 114, 245

\bibitem[{{Tielens} {et~al.}(1991){Tielens}, {Tokunaga}, {Geballe}, \&
  {Baas}}]{Tielens91}
{Tielens}, A.~G.~G.~M., {Tokunaga}, A.~T., {Geballe}, T.~R., \& {Baas}, F.
  1991, \apj, 381, 181

\bibitem[{{van der Tak} {et~al.}(2000{\natexlab{a}}){van der Tak}, {van
  Dishoeck}, \& {Caselli}}]{vanderTak00b}
{van der Tak}, F.~F.~S., {van Dishoeck}, E.~F., \& {Caselli}, P.
  2000{\natexlab{a}}, \aap, 361, 327

\bibitem[{{van der Tak} {et~al.}(2000{\natexlab{b}}){van der Tak}, {van
  Dishoeck}, {Evans}, \& {Blake}}]{vanderTak00a}
{van der Tak}, F.~F.~S., {van Dishoeck}, E.~F., {Evans}, N.~J., \& {Blake},
  G.~A. 2000{\natexlab{b}}, \apj, 537, 283

\bibitem[{{van Dishoeck} {et~al.}(1995){van Dishoeck}, {Blake}, {Jansen}, \&
  {Groesbeck}}]{vanDishoeck95}
{van Dishoeck}, E.~F., {Blake}, G.~A., {Jansen}, D.~J., \& {Groesbeck}, T.~D.
  1995, \apj, 447, 760+

\end{thebibliography}
\bibliographystyle{aa} 

\Online

\begin{table*}
\begin{center}
  \caption{Lines intensities and width for methanol transitions in
    NGC1333-IRAS4A.}
  \label{tab:flux_iras4a}
  \begin{tabular}{l l l l l l l}
    
    \hline
    \hline
    Transition & E$\mathrm{up}$ & $\nu$ & $\int{T_\mathrm{mb} \, dv}$ & $\Delta v$  & Telescope & HPBW \\       
    & (K) & (GHz) & ($\kkms$) & ($\kms$) && ($\arcsec$)\\     
    \hline

    5$_0$-4$_0$ E$^+$              &  48.0 & 241.700 & 1.70 $\pm$ 0.41 & 2.7 $\pm$ 0.4            & IRAM-30m & 10 \\
    5$_1$-4$_1$ E$^-$              &  40.5 & 241.767 & 4.63 $\pm$ 0.87 & 2.9 $\pm$ 0.2            & IRAM-30m & 10 \\
    5$_0$-4$_0$ A$^+$              &  34.8 & 241.791 & 7.46 $\pm$ 1.40 & 3.7 $\pm$ 0.2            & IRAM-30m & 10 \\
    5$_4$-4$_4$ A$^\pm$            & 115.2 & 241.806 & $<$ 0.21        & ...                      & IRAM-30m & 10 \\
    5$_4$-4$_4$ E$^-$              & 122.8 & 241.813 & $<$ 0.21        & ...                      & IRAM-30m & 10 \\
    5$_4$-4$_4$ E$^+$              & 135.8 & 241.829 & $<$ 0.21        & ...                      & IRAM-30m & 10 \\
    5$_3$-4$_3$ A$^\pm$            &  84.7 & 241.832 & $<$ 0.21        & ...                      & IRAM-30m & 10 \\
    5$_2$-4$_2$ A$^-$\tabnote{a}   &  72.6 & 241.842 & $<$ 0.21        & ...                      & IRAM-30m & 10 \\
    5$_3$-4$_3$ E$^-$              &  97.6 & 241.852 & $<$ 0.21        & ...                      & IRAM-30m & 10 \\
    5$_1$-4$_1$ E$^+$              &  55.9 & 241.879 & $<$ 0.21        & ...                      & IRAM-30m & 10 \\
    5$_2$-4$_2$ A$^+$              &  72.6 & 241.887 & $<$ 0.21        & ...                      & IRAM-30m & 10 \\
    7$_1$-6$_1$ E$^-$              &  70.6 & 338.345 & 6.71 $\pm$ 2.08 & 8.1 $\pm$ 0.1\tabnote{d} & JCMT & 13 \\
    7$_0$-6$_0$ A$^+$\tabnote{b}   &  65.1 & 338.409 & 7.95 $\pm$ 2.45 & 8.0 $\pm$ 0.8\tabnote{d} & JCMT & 13 \\
    7$_6$-6$_6$ E$^-$              & 254.7 & 338.431 & $<$ 0.06        & ...                      & JCMT & 13 \\
    7$_6$-6$_6$ A$^\pm$            & 258.9 & 338.442 & $<$ 0.06        & ...                      & JCMT & 13 \\
    7$_5$-6$_5$ E$^-$              & 189.2 & 338.457 & 0.03 $\pm$ 0.04 & 1.1 $\pm$ 1.3            & JCMT & 13 \\
    7$_5$-6$_5$ E$^+$              & 206.1 & 338.475 & $<$ 0.06        & ...                      & JCMT & 13 \\
    7$_5$-6$_5$ A$^\pm$            & 203.0 & 338.486 & $<$ 0.06        & ...                      & JCMT & 13 \\
    7$_4$-6$_4$ E$^-$              & 153.1 & 338.504 & $<$ 0.06        & ...                      & JCMT & 13 \\
    7$_2$-6$_2$ A$^-$\tabnote{c}   & 102.8 & 338.513 & 0.46 $\pm$ 0.22 & 6.9 $\pm$ 1.4\tabnote{d} & JCMT & 13 \\
    7$_4$-6$_4$ E$^+$              & 166.0 & 338.530 & $<$ 0.06        & ...                      & JCMT & 13 \\
    7$_3$-6$_3$ A$^\pm$            & 114.9 & 338.541 & 0.75 $\pm$ 0.27 & 7.4 $\pm$ 0.5\tabnote{d} & JCMT & 13 \\
    7$_3$-6$_3$ E$^-$              & 127.9 & 338.560 & $<$ 0.06        & ...                      & JCMT & 13 \\
    7$_3$-6$_3$ E$^+$              & 112.8 & 338.583 & 0.14 $\pm$ 0.07 & 3.1 $\pm$ 0.9\tabnote{d} & JCMT & 13 \\
    7$_1$-6$_1$ E$^+$              &  86.1 & 338.615 & 1.40 $\pm$ 0.47 & 7.6 $\pm$ 0.3\tabnote{d} & JCMT & 13 \\
    7$_2$-6$_2$ A$^+$              & 102.8 & 338.640 & 0.29 $\pm$ 0.13 & 6.5 $\pm$ 1.1\tabnote{d} & JCMT & 13 \\
    
    \hline
 
  \end{tabular}
\end{center}

\tabnote{a} Blended with 5$_3$-4$_3$ E$^+$.\\
\tabnote{b} Blended with 7$_6$-6$_6$ E$^+$.\\
\tabnote{c} Blended with 7$_4$-6$_4$ A$^\pm$.\\
\tabnote{d} Uncertain flux because of outflow contamination\\

\end{table*}

%%% Local Variables: 
%%% mode: latex
%%% TeX-master: "/Users/bastien/Methanol/ArticleAA/aa_ch3oh"
%%% End: 

\begin{table*}
\begin{center}
  \caption{As in Table \ref{tab:flux_iras4a} for NGC1333-IRAS4B.}
  \label{tab:flux_iras4b}
  \begin{tabular}{l l l l l l l}
    
    \hline
    \hline
    Transition & E$\mathrm{up}$ & $\nu$ & $\int{T_\mathrm{mb} \, dv}$ & $\Delta v$  & Telescope & HPBW \\       
    & (K) & (GHz) & ($\kkms$) & ($\kms$) && ($\arcsec$)\\     
    \hline
    
    5$_0$-4$_0$ E$^+$              &  48.0 & 241.700 & 1.56 $\pm$ 0.50 & 1.2 $\pm$ 0.4 & IRAM-30m & 10 \\ 
    5$_1$-4$_1$ E$^-$              &  40.5 & 241.767 & 2.39 $\pm$ 0.53 & 1.1 $\pm$ 0.1 & IRAM-30m & 10 \\ 
    5$_0$-4$_0$ A$^+$              &  34.8 & 241.791 & 3.74 $\pm$ 0.77 & 1.3 $\pm$ 0.1 & IRAM-30m & 10 \\ 
    5$_4$-4$_4$ A$^\pm$            & 115.2 & 241.806 & $<$ 0.30        & ...           & IRAM-30m & 10 \\ 
    5$_4$-4$_4$ E$^-$              & 122.8 & 241.813 & $<$ 0.30        & ...           & IRAM-30m & 10 \\ 
    5$_4$-4$_4$ E$^+$              & 135.8 & 241.829 & $<$ 0.30        & ...           & IRAM-30m & 10 \\ 
    5$_3$-4$_3$ A$^\pm$            &  84.7 & 241.832 & $<$ 0.30        & ...           & IRAM-30m & 10 \\ 
    5$_2$-4$_2$ A$^-$\tabnote{a}   &  76.8 & 241.842 & $<$ 0.30        & ...           & IRAM-30m & 10 \\ 
    5$_3$-4$_3$ E$^-$              &  97.6 & 241.852 & $<$ 0.30        & ...           & IRAM-30m & 10 \\ 
    5$_1$-4$_1$ E$^+$              &  55.9 & 241.879 & $<$ 0.30        & ...           & IRAM-30m & 10 \\ 
    5$_2$-4$_2$ A$^+$              &  72.6 & 241.887 & $<$ 0.30        & ...           & IRAM-30m & 10 \\ 
    7$_1$-6$_1$ E$^-$              &  70.6 & 338.345 & 5.27 $\pm$ 1.66 & 3.7 $\pm$ 0.1 & JCMT & 13 \\     
    7$_0$-6$_0$ A$^+$\tabnote{b}   &  65.1 & 338.409 & 5.78 $\pm$ 1.81 & 3.4 $\pm$ 0.6 & JCMT & 13 \\     
    7$_6$-6$_6$ E$^-$              & 254.7 & 338.431 & $<$ 0.08        & ...           & JCMT & 13 \\     
    7$_6$-6$_6$ A$^\pm$            & 258.9 & 338.442 & $<$ 0.08        & ...           & JCMT & 13 \\     
    7$_5$-6$_5$ E$^-$              & 189.2 & 338.457 & $<$ 0.08        & ...           & JCMT & 13 \\     
    7$_5$-6$_5$ E$^+$              & 206.1 & 338.475 & $<$ 0.08        & ...           & JCMT & 13 \\     
    7$_5$-6$_5$ A$^\pm$            & 203.0 & 338.486 & $<$ 0.08        & ...           & JCMT & 13 \\     
    7$_4$-6$_4$ E$^-$              & 153.1 & 338.504 & $<$ 0.08        & ...           & JCMT & 13 \\     
    7$_2$-6$_2$ A$^-$\tabnote{c}   & 131.2 & 338.513 & 0.76 $\pm$ 0.29 & 4.7 $\pm$ 0.5 & JCMT & 13 \\     
    7$_4$-6$_4$ E$^+$              & 166.0 & 338.530 & $<$ 0.08        & ...           & JCMT & 13 \\     
    7$_3$-6$_3$ A$^\pm$            & 114.9 & 338.541 & 1.29 $\pm$ 0.47 & 5.3 $\pm$ 0.4 & JCMT & 13 \\     
    7$_3$-6$_3$ E$^-$              & 127.9 & 338.560 & $<$ 0.08        & ...           & JCMT & 13 \\     
    7$_3$-6$_3$ E$^+$              & 112.8 & 338.583 & 0.30 $\pm$ 0.19 & 3.0 $\pm$ 1.0 & JCMT & 13 \\     
    7$_1$-6$_1$ E$^+$              &  86.1 & 338.615 & 1.60 $\pm$ 0.61 & 4.2 $\pm$ 0.4 & JCMT & 13 \\     
    7$_2$-6$_2$ A$^+$              & 102.8 & 338.640 & 0.52 $\pm$ 0.27 & 4.0 $\pm$ 1.1 & JCMT & 13 \\     
    
    \hline
    
  \end{tabular}
\end{center}

\tabnote{a} Blended with 5$_3$-4$_3$ E$^+$.\\
\tabnote{b} Blended with 7$_6$-6$_6$ E$^+$.\\
\tabnote{c} Blended with 7$_4$-6$_4$ A$^\pm$.\\

\end{table*}

%%% Local Variables: 
%%% mode: latex
%%% TeX-master: "/Users/bastien/Methanol/ArticleAA/aa_ch3oh"
%%% End: 

\begin{table*}
\begin{center}
  \caption{As in Table \ref{tab:flux_iras4a} for NGC1333-IRAS2.}
  \label{tab:flux_iras2}
  \begin{tabular}{l l l l l l l}
    
    \hline
    \hline
    Transition & E$\mathrm{up}$ & $\nu$ & $\int{T_\mathrm{mb} \, dv}$ & $\Delta v$  & Telescope & HPBW \\       
    & (K) & (GHz) & ($\kkms$) & ($\kms$) && ($\arcsec$)\\     
    \hline

    5$_0$-4$_0$ E$^+$              &  48.0 & 241.700 & 0.89 $\pm$ 0.34 & 2.9 $\pm$ 0.8            & IRAM-30m & 10 \\   % 503-403
    5$_1$-4$_1$ E$^-$              &  40.5 & 241.767 & 1.67 $\pm$ 0.41 & 2.4 $\pm$ 0.3            & IRAM-30m & 10 \\   % 514-411
    5$_0$-4$_0$ A$^+$              &  34.8 & 241.791 & 1.74 $\pm$ 0.40 & 2.0 $\pm$ 0.2            & IRAM-30m & 10 \\   % 501-401
    5$_4$-4$_4$ A$^\pm$            & 115.2 & 241.806 & 0.54 $\pm$ 0.20 & 3.5 $\pm$ 1.0            & IRAM-30m & 10 \\   % 542-442
    5$_4$-4$_4$ E$^-$              & 122.8 & 241.813 & 0.47 $\pm$ 0.19 & 3.3 $\pm$ 1.0            & IRAM-30m & 10 \\   % 544-444
    5$_4$-4$_4$ E$^+$              & 135.8 & 241.829 & 0.42 $\pm$ 0.22 & 3.4 $\pm$ 1.3            & IRAM-30m & 10 \\   % 543-443
    5$_3$-4$_3$ A$^\pm$            &  84.7 & 241.832 & 0.74 $\pm$ 0.27 & 3.2 $\pm$ 0.8            & IRAM-30m & 10 \\   % 532-432
    5$_2$-4$_2$ A$^-$\tabnote{a}   &  76.8 & 241.842 & 0.75 $\pm$ 0.22 & 3.6 $\pm$ 0.5            & IRAM-30m & 10 \\   % 533-434
    5$_3$-4$_3$ E$^-$              &  97.6 & 241.852 & 0.39 $\pm$ 0.16 & 2.8 $\pm$ 0.9            & IRAM-30m & 10 \\   % 534-434
    5$_1$-4$_1$ E$^+$              &  55.9 & 241.879 & 1.00 $\pm$ 0.26 & 3.5 $\pm$ 0.5            & IRAM-30m & 10 \\   % 513-413
    5$_2$-4$_2$ A$^+$              &  72.6 & 241.887 & 0.63 $\pm$ 0.18 & 2.7 $\pm$ 0.5            & IRAM-30m & 10 \\   % 511-411
    7$_1$-6$_1$ E$^-$              &  70.6 & 338.345 & 1.59 $\pm$ 0.59 & 5.9 $\pm$ 0.5\tabnote{d} & JCMT & 13 \\       % 714-614
    7$_0$-6$_0$ A$^+$\tabnote{b}   &  65.1 & 338.409 & 1.89 $\pm$ 0.68 & 6.5 $\pm$ 0.5\tabnote{d} & JCMT & 13 \\       % 701-601
    7$_6$-6$_6$ E$^-$              & 254.7 & 338.431 & 0.27 $\pm$ 0.14 & 3.0 $\pm$ 0.8            & JCMT & 13 \\       % 764-664 
    7$_6$-6$_6$ A$^\pm$            & 258.9 & 338.442 & 0.13 $\pm$ 0.12 & 3.2 $\pm$ 2.4            & JCMT & 13 \\       % 762-662 
    7$_5$-6$_5$ E$^-$              & 189.2 & 338.457 & 0.27 $\pm$ 0.19 & 3.1 $\pm$ 2.0            & JCMT & 13 \\       % 754-654 
    7$_5$-6$_5$ E$^+$              & 206.1 & 338.475 & 0.27 $\pm$ 0.18 & 2.7 $\pm$ 1.0            & JCMT & 13 \\       % 753-653
    7$_5$-6$_5$ A$^\pm$            & 203.0 & 338.486 & 0.29 $\pm$ 0.20 & 3.0 $\pm$ 1.7            & JCMT & 13 \\       % 752-652
    7$_4$-6$_4$ E$^-$              & 153.1 & 338.504 & 0.37 $\pm$ 0.22 & 3.4 $\pm$ 1.3            & JCMT & 13 \\       % 744-644
    7$_2$-6$_2$ A$^-$\tabnote{c}   & 131.2 & 338.513 & 0.40 $\pm$ 0.23 & 3.2 $\pm$ 1.3            & JCMT & 13 \\       % 741-641
    7$_4$-6$_4$ E$^+$              & 166.0 & 338.530 & 0.43 $\pm$ 0.24 & 3.2 $\pm$ 1.1            & JCMT & 13 \\       % 743-643
    7$_3$-6$_3$ A$^\pm$            & 114.9 & 338.541 & 0.71 $\pm$ 0.34 & 4.8 $\pm$ 1.0            & JCMT & 13 \\       % 732-632
    7$_3$-6$_3$ E$^-$              & 127.9 & 338.560 & 0.33 $\pm$ 0.20 & 3.2 $\pm$ 0.8            & JCMT & 13 \\       % 734-634 
    7$_3$-6$_3$ E$^+$              & 112.8 & 338.583 & 0.41 $\pm$ 0.22 & 3.1 $\pm$ 0.9            & JCMT & 13 \\       % 733-633
    7$_1$-6$_1$ E$^+$              &  86.1 & 338.615 & 0.67 $\pm$ 0.33 & 4.2 $\pm$ 1.1            & JCMT & 13 \\       % 713-613
    7$_2$-6$_2$ A$^+$              & 102.8 & 338.640 & 0.37 $\pm$ 0.19 & 2.6 $\pm$ 0.7            & JCMT & 13 \\       % 721-621
    
    \hline

  \end{tabular}
\end{center}

\tabnote{a} Blended with 5$_3$-4$_3$ E$^+$.\\
\tabnote{b} Blended with 7$_6$-6$_6$ E$^+$.\\
\tabnote{c} Blended with 7$_4$-6$_4$ A$^\pm$.\\
\tabnote{d} Uncertain flux because of outflow contamination\\

\end{table*}

%%% Local Variables: 
%%% mode: latex
%%% TeX-master: "/Users/bastien/Methanol/ArticleAA/aa_ch3oh"
%%% End: 

\begin{table*}
\begin{center}
  \caption{As in Table \ref{tab:flux_iras4a} for L1448-MM.}
  \label{tab:flux_l1448mm}
  \begin{tabular}{l l l l l l l}
    
    \hline
    \hline
    Transition & E$\mathrm{up}$ & $\nu$ & $\int{T_\mathrm{mb} \, dv}$ & $\Delta v$  & Telescope & HPBW \\       
    & (K) & (GHz) & ($\kkms$) & ($\kms$) && ($\arcsec$)\\     
    \hline\

    5$_0$-4$_0$ E$^+$              &  48.0 & 241.700 & 0.30 $\pm$ 0.15 & 1.3 $\pm$ 0.1 & IRAM-30m & 10 \\
    5$_1$-4$_1$ E$^-$              &  40.5 & 241.767 & 0.70 $\pm$ 0.14 & 1.2 $\pm$ 0.1 & IRAM-30m & 10 \\
    5$_0$-4$_0$ A$^+$              &  34.8 & 241.791 & 0.93 $\pm$ 0.17 & 1.1 $\pm$ 0.1 & IRAM-30m & 10 \\
    5$_4$-4$_4$ A$^\pm$            & 115.2 & 241.806 & $<$ 0.05        & ...           & IRAM-30m & 10 \\
    5$_4$-4$_4$ E$^-$              & 122.8 & 241.813 & $<$ 0.05        & ...           & IRAM-30m & 10 \\
    5$_4$-4$_4$ E$^+$              & 135.8 & 241.829 & $<$ 0.05        & ...           & IRAM-30m & 10 \\
    5$_3$-4$_3$ A$^\pm$            &  84.7 & 241.832 & $<$ 0.05        & ...           & IRAM-30m & 10 \\
    5$_2$-4$_2$ A$^-$\tabnote{a}   &  76.8 & 241.842 & 0.26 $\pm$ 0.07 & 3.5 $\pm$ 0.6 & IRAM-30m & 10 \\
    5$_3$-4$_3$ E$^-$              &  97.6 & 241.852 & 0.25 $\pm$ 0.07 & 2.0 $\pm$ 0.4 & IRAM-30m & 10 \\
    5$_1$-4$_1$ E$^+$              &  55.9 & 241.879 & 0.32 $\pm$ 0.08 & 1.8 $\pm$ 0.3 & IRAM-30m & 10 \\
    5$_2$-4$_2$ A$^+$              &  72.6 & 241.887 & $<$ 0.05        & ...           & IRAM-30m & 10 \\
    7$_1$-6$_1$ E$^-$              &  70.6 & 338.345 & 0.60 $\pm$ 0.36 & 7.5 $\pm$ 2.5 & JCMT & 13 \\
    7$_0$-6$_0$ A$^+$\tabnote{b}   &  65.1 & 338.409 & 0.54 $\pm$ 0.24 & 5.9 $\pm$ 1.0 & JCMT & 13 \\    
    7$_6$-6$_6$ E$^-$              & 254.7 & 338.431 & $<$ 0.11        & ...           & JCMT & 13 \\    
    7$_6$-6$_6$ A$^\pm$            & 258.9 & 338.442 & $<$ 0.11        & ...           & JCMT & 13 \\    
    7$_5$-6$_5$ E$^-$              & 189.2 & 338.457 & $<$ 0.11        & ...           & JCMT & 13 \\    
    7$_5$-6$_5$ E$^+$              & 206.1 & 338.475 & $<$ 0.11        & ...           & JCMT & 13 \\    
    7$_5$-6$_5$ A$^\pm$            & 203.0 & 338.486 & $<$ 0.11        & ...           & JCMT & 13 \\    
    7$_4$-6$_4$ E$^-$              & 153.1 & 338.504 & $<$ 0.11        & ...           & JCMT & 13 \\    
    7$_2$-6$_2$ A$^-$\tabnote{c}   & 131.2 & 338.513 & $<$ 0.11        & ...           & JCMT & 13 \\    
    7$_4$-6$_4$ E$^+$              & 166.0 & 338.530 & $<$ 0.11        & ...           & JCMT & 13 \\    
    7$_3$-6$_3$ A$^\pm$            & 114.9 & 338.541 & $<$ 0.11        & ...           & JCMT & 13 \\    
    7$_3$-6$_3$ E$^-$              & 127.9 & 338.560 & $<$ 0.11        & ...           & JCMT & 13 \\    
    7$_3$-6$_3$ E$^+$              & 112.8 & 338.583 & $<$ 0.11        & ...           & JCMT & 13 \\    
    7$_1$-6$_1$ E$^+$              &  86.1 & 338.615 & 0.30 $\pm$ 0.19 & 7.9 $\pm$ 2.7 & JCMT & 13 \\    
    7$_2$-6$_2$ A$^+$              & 102.8 & 338.640 & $<$ 0.11        & ...           & JCMT & 13 \\ 

    \hline
    
  \end{tabular}
\end{center}

\tabnote{a} Blended with 5$_3$-4$_3$ E$^+$.\\
\tabnote{b} Blended with 7$_6$-6$_6$ E$^+$.\\
\tabnote{c} Blended with 7$_4$-6$_4$ A$^\pm$.\\

\end{table*}

%%% Local Variables: 
%%% mode: latex
%%% TeX-master: "/Users/bastien/Methanol/ArticleAA/aa_ch3oh"
%%% End: 

\begin{table*}
\begin{center}
  \caption{As in Table \ref{tab:flux_iras4a} for L1448-N.}
  \label{tab:flux_l1448n}
  \begin{tabular}{l l l l l l l}
    
    \hline
    \hline
    Transition & E$\mathrm{up}$ & $\nu$ & $\int{T_\mathrm{mb} \, dv}$ & $\Delta v$  & Telescope & HPBW \\       
    & (K) & (GHz) & ($\kkms$) & ($\kms$) && ($\arcsec$)\\     
    \hline
    
    5$_0$-4$_0$ E$^+$              &  48.0 & 241.700 & 0.39 $\pm$ 0.15 & 1.2 $\pm$ 0.4 & IRAM-30m & 10 \\ 
    5$_1$-4$_1$ E$^-$              &  40.5 & 241.767 & 1.00 $\pm$ 0.20 & 1.1 $\pm$ 0.1 & IRAM-30m & 10 \\ 
    5$_0$-4$_0$ A$^+$              &  34.8 & 241.791 & 1.33 $\pm$ 0.25 & 1.3 $\pm$ 0.1 & IRAM-30m & 10 \\ 
    5$_4$-4$_4$ A$^\pm$            & 115.2 & 241.806 & $<$ 0.05        & ...           & IRAM-30m & 10 \\ 
    5$_4$-4$_4$ E$^-$              & 122.8 & 241.813 & $<$ 0.05        & ...           & IRAM-30m & 10 \\ 
    5$_4$-4$_4$ E$^+$              & 135.8 & 241.829 & $<$ 0.05        & ...           & IRAM-30m & 10 \\ 
    5$_3$-4$_3$ A$^\pm$            &  84.7 & 241.832 & $<$ 0.05        & ...           & IRAM-30m & 10 \\ 
    5$_2$-4$_2$ A$^-$\tabnote{a}   &  76.8 & 241.842 & $<$ 0.05        & ...           & IRAM-30m & 10 \\ 
    5$_3$-4$_3$ E$^-$              &  97.6 & 241.852 & $<$ 0.05        & ...           & IRAM-30m & 10 \\ 
    5$_1$-4$_1$ E$^+$              &  55.9 & 241.879 & 0.26 $\pm$ 0.09 & 2.6 $\pm$ 0.6 & IRAM-30m & 10 \\ 
    5$_2$-4$_2$ A$^+$              &  72.6 & 241.887 & $<$ 0.05        & ...           & IRAM-30m & 10 \\ 
    7$_1$-6$_1$ E$^-$              &  70.6 & 338.345 & 0.38 $\pm$ 0.18 & 1.8 $\pm$ 0.3 & JCMT & 13 \\     
    7$_0$-6$_0$ A$^+$\tabnote{b}   &  65.1 & 338.409 & 0.48 $\pm$ 0.21 & 2.0 $\pm$ 0.3 & JCMT & 13 \\     
    7$_6$-6$_6$ E$^-$              & 254.7 & 338.431 & $<$ 0.06        & ...           & JCMT & 13 \\     
    7$_6$-6$_6$ A$^\pm$            & 258.9 & 338.442 & $<$ 0.06        & ...           & JCMT & 13 \\     
    7$_5$-6$_5$ E$^-$              & 189.2 & 338.457 & $<$ 0.06        & ...           & JCMT & 13 \\     
    7$_5$-6$_5$ E$^+$              & 206.1 & 338.475 & $<$ 0.06        & ...           & JCMT & 13 \\     
    7$_5$-6$_5$ A$^\pm$            & 203.0 & 338.486 & $<$ 0.06        & ...           & JCMT & 13 \\     
    7$_4$-6$_4$ E$^-$              & 153.1 & 338.504 & $<$ 0.06        & ...           & JCMT & 13 \\     
    7$_2$-6$_2$ A$^-$\tabnote{c}   & 131.2 & 338.513 & $<$ 0.06        & ...           & JCMT & 13 \\     
    7$_4$-6$_4$ E$^+$              & 166.0 & 338.530 & $<$ 0.06        & ...           & JCMT & 13 \\     
    7$_3$-6$_3$ A$^\pm$            & 114.9 & 338.541 & 0.11 $\pm$ 0.10 & 2.8 $\pm$ 1.6 & JCMT & 13 \\     
    7$_3$-6$_3$ E$^-$              & 127.9 & 338.560 & $<$ 0.06        & ...           & JCMT & 13 \\     
    7$_3$-6$_3$ E$^+$              & 112.8 & 338.583 & $<$ 0.06        & ...           & JCMT & 13 \\     
    7$_1$-6$_1$ E$^+$              &  86.1 & 338.615 & $<$ 0.06        & ...           & JCMT & 13 \\     
    7$_2$-6$_2$ A$^+$              & 102.8 & 338.640 & $<$ 0.06        & ...           & JCMT & 13 \\     

    \hline

  \end{tabular}
\end{center}

\tabnote{a} Blended with 5$_3$-4$_3$ E$^+$.\\
\tabnote{b} Blended with 7$_6$-6$_6$ E$^+$.\\
\tabnote{c} Blended with 7$_4$-6$_4$ A$^\pm$.\\

\end{table*}

%%% Local Variables: 
%%% mode: latex
%%% TeX-master: "/Users/bastien/Methanol/ArticleAA/aa_ch3oh"
%%% End: 

\begin{table*}
\begin{center}
  \caption{As in Table \ref{tab:flux_iras4a} for L1157-MM.}
  \label{tab:flux_l1157mm}
  \begin{tabular}{l l l l l l l}
    
    \hline
    \hline
    Transition & E$\mathrm{up}$ & $\nu$ & $\int{T_\mathrm{mb} \, dv}$ & $\Delta v$  & Telescope & HPBW \\       
    & (K) & (GHz) & ($\kkms$) & ($\kms$) && ($\arcsec$) \\     
    \hline
    
    5$_0$-4$_0$ E$^+$              &  48.0 & 241.700 & $<$ 0.07        & ...           & IRAM-30m & 10 \\
    5$_1$-4$_1$ E$^-$              &  40.5 & 241.767 & 0.51 $\pm$ 0.13 & 1.2 $\pm$ 0.2 & IRAM-30m & 10 \\
    5$_0$-4$_0$ A$^+$              &  34.8 & 241.791 & 0.54 $\pm$ 0.13 & 1.2 $\pm$ 0.2 & IRAM-30m & 10 \\
    5$_4$-4$_4$ A$^\pm$            & 115.2 & 241.806 & $<$ 0.07        & ...           & IRAM-30m & 10 \\
    5$_4$-4$_4$ E$^-$              & 122.8 & 241.813 & $<$ 0.07        & ...           & IRAM-30m & 10 \\
    5$_4$-4$_4$ E$^+$              & 135.8 & 241.829 & $<$ 0.07        & ...           & IRAM-30m & 10 \\
    5$_3$-4$_3$ A$^\pm$            &  84.7 & 241.832 & $<$ 0.07        & ...           & IRAM-30m & 10 \\
    5$_2$-4$_2$ A$^-$\tabnote{a}   &  76.8 & 241.842 & $<$ 0.07        & ...           & IRAM-30m & 10 \\
    5$_3$-4$_3$ E$^-$              &  97.6 & 241.852 & $<$ 0.07        & ...           & IRAM-30m & 10 \\
    5$_1$-4$_1$ E$^+$              &  55.9 & 241.879 & $<$ 0.07        & ...           & IRAM-30m & 10 \\
    5$_2$-4$_2$ A$^+$              &  72.6 & 241.887 & $<$ 0.07        & ...           & IRAM-30m & 10 \\
    7$_1$-6$_1$ E$^-$              &  70.6 & 338.345 & $<$ 0.21        & ...           & JCMT & 13 \\     
    7$_0$-6$_0$ A$^+$\tabnote{b}   &  65.1 & 338.409 & 0.41 $\pm$ 0.27 & 7.8 $\pm$ 3.7 & JCMT & 13 \\     
    7$_6$-6$_6$ E$^-$              & 254.7 & 338.431 & $<$ 0.21        & ...           & JCMT & 13 \\     
    7$_6$-6$_6$ A$^\pm$            & 258.9 & 338.442 & $<$ 0.21        & ...           & JCMT & 13 \\     
    7$_5$-6$_5$ E$^-$              & 189.2 & 338.457 & $<$ 0.21        & ...           & JCMT & 13 \\     
    7$_5$-6$_5$ E$^+$              & 206.1 & 338.475 & $<$ 0.21        & ...           & JCMT & 13 \\     
    7$_5$-6$_5$ A$^\pm$            & 203.0 & 338.486 & $<$ 0.21        & ...           & JCMT & 13 \\     
    7$_4$-6$_4$ E$^-$              & 153.1 & 338.504 & $<$ 0.21        & ...           & JCMT & 13 \\     
    7$_2$-6$_2$ A$^-$\tabnote{c}   & 131.2 & 338.513 & $<$ 0.21        & ...           & JCMT & 13 \\     
    7$_4$-6$_4$ E$^+$              & 166.0 & 338.530 & $<$ 0.21        & ...           & JCMT & 13 \\     
    7$_3$-6$_3$ A$^\pm$            & 114.9 & 338.541 & $<$ 0.21        & ...           & JCMT & 13 \\     
    7$_3$-6$_3$ E$^-$              & 127.9 & 338.560 & $<$ 0.21        & ...           & JCMT & 13 \\     
    7$_3$-6$_3$ E$^+$              & 112.8 & 338.583 & $<$ 0.21        & ...           & JCMT & 13 \\     
    7$_1$-6$_1$ E$^+$              &  86.1 & 338.615 & $<$ 0.21        & ...           & JCMT & 13 \\     
    7$_2$-6$_2$ A$^+$              & 102.8 & 338.640 & $<$ 0.21        & ...           & JCMT & 13 \\    
    
    \hline
    
  \end{tabular}
\end{center}

\tabnote{a} Blended with 5$_3$-4$_3$ E$^+$.\\
\tabnote{b} Blended with 7$_6$-6$_6$ E$^+$.\\
\tabnote{c} Blended with 7$_4$-6$_4$ A$^\pm$.\\

\end{table*}

%%% Local Variables: 
%%% mode: latex
%%% TeX-master: "/Users/bastien/Methanol/ArticleAA/aa_ch3oh"
%%% End: 

\begin{table*}
\begin{center}
  \caption{As in Table \ref{tab:flux_iras4a} for IRAS16293-2422.}
  \label{tab:flux_iras16293}
  \begin{tabular}{l l l l l l l}
    
    \hline
    \hline
    Transition & E$\mathrm{up}$ & $\nu$ & $\int{T_\mathrm{mb} \, dv}$  & $\Delta v$  & Telescope & HPBW \\       
    & (K) & (GHz) & ($\kkms$) & ($\kms$) && ($\arcsec$) \\     
    \hline

    5$_0$-4$_0$ E$^+$            &  48.0 & 241.700 & 2.67 $\pm$ 0.83 & 4.5 $\pm$ 1.0    & JCMT & 20 \\
    5$_1$-4$_1$ E$^-$            &  40.5 & 241.767 & 4.36 $\pm$ 0.97 & 3.3 $\pm$ 0.3    & JCMT & 20 \\
    5$_0$-4$_0$ A$^+$            &  34.8 & 241.791 & 4.91 $\pm$ 1.03 & 3.2 $\pm$ 0.2    & JCMT & 20 \\
    5$_4$-4$_4$ A$^\pm$          & 115.2 & 241.806 & 0.72 $\pm$ 0.46 & fixed\tabnote{a} & JCMT & 20 \\
    5$_4$-4$_4$ E$^-$            & 122.8 & 241.813 & $<$ 0.10        & ...              & JCMT & 20 \\
    5$_4$-4$_4$ E$^+$            & 135.8 & 241.829 & $<$ 0.10        & ...              & JCMT & 20 \\
    5$_3$-4$_3$ A$^\pm$          &  84.7 & 241.832 & 1.99 $\pm$ 0.75 & 5.3 $\pm$ 1.4    & JCMT & 20 \\
    5$_2$-4$_2$ A$^-$\tabnote{b} &  72.6 & 241.842 & 1.97 $\pm$ 0.77 & 6.3 $\pm$ 1.7    & JCMT & 20 \\
    5$_3$-4$_3$ E$^-$            &  97.6 & 241.852 & 0.61 $\pm$ 0.44 & fixed\tabnote{a} & JCMT & 20 \\
    5$_1$-4$_1$ E$^+$            &  55.9 & 241.879 & 2.29 $\pm$ 0.78 & 4.9 $\pm$ 1.2    & JCMT & 20 \\
    5$_2$-4$_2$ A$^+$            &  72.6 & 241.887 & 1.67 $\pm$ 0.82 & 7.8 $\pm$ 3.0    & JCMT & 20 \\
    5$_2$-4$_2$ E$^-$            &  60.8 & 241.904 & 3.28 $\pm$ 0.91 & 4.6 $\pm$ 0.7    & JCMT & 20 \\
    
    \hline

  \end{tabular}
\end{center}

\tabnote{a} Gaussian fit with a fixed line width of 5.0 km.s$^{-1}$.\\
\tabnote{b} Blended with 5$_3$-4$_3$ E$^+$.\\

\end{table*}

%%% Local Variables: 
%%% mode: latex
%%% TeX-master: "/Users/bastien/Methanol/ArticleAA/aa_ch3oh"
%%% End: 

\appendix
\onecolumn
% Labeling of energy levels and selection rules for
% interstellar methanol
% Alexandre Faure, 4 June 2004

\section{Spectral designation}
\label{sec:spectral-designation}

Methanol (CH$_3$OH) is a slightly asymmetric rotor with hindered
internal rotation of the methyl (CH$_3$) group. The first excited
state of the torsional vibration is about 200~cm$^{-1}$ above the
ground state, low enough to be populated in some particular
interstellar conditions \citep{Menten86}. Energy levels are labelled
by the total angular momentum $J$, its projection along the symmetry
$a$-axis $k$ or $K=|k|$, the torsional vibration quantum number $v_t$
and the torsional sublevel quantum number $\sigma$. The threefold
symmetric torsional barrier in methanol indeed leads to the existence
of three torsional symmetry states, usually designated $A$, $E_1$ and
$E_2$ corresponding to $\sigma=0, +1, -1$, respectively
\citep{Lees68}. For the $E$ species, $k\geq 0$ states are degenerate
with the $E_2$ $k\leq 0$ states, and vice versa. It is thus simplest
to refer to these states as doubly degenerate states of $E$ symmetry
with positive and negative values for $k$ \citep{Lees73}. Levels of
the $A$ species are torsionally doubly degenerate but this degeneracy
is lifted by the slight asymmetry of the molecule. These doublets may
be labelled by the $C_{3v}$ symmetry group notation $A_1$ and $A_2$
but the most common convention is tu use the labels $A^+$ and
$A^-$. Full details about notations can be found in \citet{Lees68}. An
alternate convention for labelling the $A$ symmetry states is the use
of the standard asymmetric top notation $J_{K_aK_c}$, where $K_a$
denotes the above $K$, as done for example in the Cologne database for
molecular spectroscopy \citep{Muller01}. In this paper, the former
notation is employed.

\end{document}